\documentclass[conference, 10pt]{IEEEtran}

\IEEEoverridecommandlockouts
\usepackage{cite}
\usepackage{amsmath,amssymb,amsfonts}
\usepackage{algorithmic}
\usepackage{graphicx}
\usepackage{textcomp}
\usepackage{booktabs} 

\usepackage{subfig}
\usepackage{multirow}

\usepackage[normalem]{ulem}

\newcommand{\HWsporadic}[0]{\texttt{SP-LPPM}}
\newcommand{\HWmarkov}[0]{\texttt{MK-LPPM}}
\newcommand{\LH}[0]{\texttt{LH}}
\newcommand{\Expo}[0]{\texttt{Exp}}
\newcommand{\LHsp}[0]{\texttt{SP-LH}}
\newcommand{\Exposp}[0]{\texttt{SP-Exp}}
\newcommand{\LHmk}[0]{\texttt{MK-LH}}
\newcommand{\Expomk}[0]{\texttt{MK-Exp}}
\newcommand{\LHbs}[0]{\texttt{PEB-LH}}
\newcommand{\Expobs}[0]{\texttt{PEB-Exp}}

\newcommand{\Qavg}[0]{\overline{\text{Q}}} 
\newcommand{\Qmax}[0]{\overline{\text{Q}}_\texttt{max}} 
\renewcommand{\P}[0]{\text{P}} 
\newcommand{\PAE}[0]{\P_{\texttt{AE}}} 

\newcommand{\dP}[1]{d_{P}(#1)} 
\newcommand{\dQ}[1]{d_{Q}(#1)} 

\newcommand{\para}[1]{\vspace{3pt}\noindent \textbf{#1}}

\usepackage{xargs}
\usepackage[colorinlistoftodos,prependcaption,textsize=tiny]{todonotes}
\newcommandx{\scomment}[2][1=]{\todo[linecolor=blue,backgroundcolor=blue!25,bordercolor=blue,#1]{#2}}
\graphicspath{{./img/}}

\newcommand{\Exp}[1]{\text{E}\{ #1 \}}
\newcommand{\Expemp}[1]{\text{E}_\texttt{emp}\{ #1 \}}

\begin{document}
 \title{Rethinking Location Privacy for\\
 Unknown Mobility Behaviors}

\author{\IEEEauthorblockN{Simon Oya}
\IEEEauthorblockA{\textit{Signal Processing in Communications Group}\\
\textit{University of Vigo}\\
Vigo, Spain\\
simonoya@gts.uvigo.es}
\and
\IEEEauthorblockN{Carmela Troncoso}
\IEEEauthorblockA{\textit{SPRING Lab}\\
\textit{EPFL}\\
Lausanne, Switzerland\\
carmela.troncoso@epfl.ch}
\and
\IEEEauthorblockN{Fernando P{\'e}rez-Gonz{\'a}lez}
\IEEEauthorblockA{\textit{Signal Processing in Communications Group}\\
\textit{University of Vigo}\\
Vigo, Spain\\
fperez@gts.uvigo.es}
}

\maketitle

\begin{abstract} 
Location Privacy-Preserving Mechanisms (LPPMs) in the literature largely consider that users' data available for training wholly characterizes their mobility patterns. Thus, they \emph{hardwire} this information in their designs and evaluate their privacy properties with these same data. In this paper, we aim to understand the impact of this decision on the level of privacy these LPPMs may offer in real life when the users' mobility data may be different from the data used in the design phase. Our results show that, in many cases, training data does not capture users' behavior accurately and, thus, the level of privacy provided by the LPPM is often overestimated. To address this gap between theory and practice, we propose to use \emph{blank-slate models} for LPPM design. Contrary to the hardwired approach, that assumes known users' behavior, blank-slate models \emph{learn} the users' behavior from the queries to the service provider. We leverage this blank-slate approach to develop a new family of LPPMs, that we call Profile Estimation-Based LPPMs. Using real data, we empirically show that our proposal outperforms optimal state-of-the-art mechanisms designed on \emph{sporadic} hardwired models. On \emph{non-sporadic} location privacy scenarios, our method is only better if the usage of the location privacy service is not continuous. It is our hope that eliminating the need to bootstrap the mechanisms with training data and ensuring that the mechanisms are lightweight and easy to compute help fostering the integration of location privacy protections in deployed systems.
\end{abstract}

\section{Introduction}

In the last decade, the research community has progressed significantly in the theoretical study and development of Location Privacy-Preserving Mechanisms (LPPMs)~\cite{Quant2011SP,OptGeoInd2014CCS,Elastic2015PETS,Semantics2016PETS}, including a number of optimal defense proposals~\cite{Opt2012CCS,herrmann2013optimal,OptGeoInd2014CCS,Elastic2015PETS,Semantics2016PETS,shokri2017privacy,chatzikokolakis2017efficient,oya2017back}. 
In practice, however, there are few LPPMs deployed, and the most popular ones are not effective at protecting the users' privacy, since they either build countermeasures that only protect against naive trilateration attacks~\cite{polakis2015s} or implement algorithms~\cite{LocationGuard} that are known to be suboptimal~\cite{chatzikokolakis2017efficient,OyaTP17}.
A potential reason for the lack of adoption and deployment of the new and effective LPPMs proposed by academia is that they require mobility traces in order to bootstrap their design.
In practice, gathering mobility data that is sufficient, up-to-date, and truly representative of a particular user's behavior is complicated. In most cases, user behavior is to some degree \emph{unknown}, and it is not clear how the latest LPPMs perform when built upon data that does not completely describe the user's mobility behavior.

Chatzikokolakis et al.~\cite{chatzikokolakis2017efficient} already claim that a fair assessment of LPPMs requires the separation between the \emph{training} dataset used for design, and the \emph{testing} dataset used for evaluation. Yet, their design strategy, as the rest of the previous works, hardwires the training mobility model into the mechanism and they do not quantify how much privacy is lost in practice when the users' mobility characteristics differs from the training data.

In this work, we aim at understanding the privacy loss associated to this discrepancy between design and deployment phases. We study both sporadic cases, where users occasionally query the Location Based Service (LBS) and thus every time their location is independent from previous LBS uses; and continuous cases, where users' actual location at a certain time depends on previously visited locations. We find that, since the design strategies in previous works hardwire the training information on the LPPMs they produce, they cannot adapt to behavioral patterns not available in the training data. We empirically show that, indeed, previous analyses overestimate the protection of the optimal LPPMs when they are evaluated on mobility profiles different from the training data. 

In response to this problem, we introduce a new design strategy that builds on what we call \emph{blank-slate models} for user mobility. Contrary to hardwired models, blank-slate models do not fix their parameters based on training data, but learn these parameters as they observe the user behavior. We take the particular case of sporadic location privacy and leverage a blank-slate model to build a new family of defenses that we call Profile Estimation-Based LPPMs (PEB-LPPMs). Like traditional LPPMs, these mechanisms are initialized with training data. However, as the user queries the LBS, they adapt their parameters. Thus, they are more adequate for those users whose behavior is not well-represented in the training data. We empirically compare PEB-LPPMs with state-of-the-art LPPMs using real data. Our evaluation confirms that PEB-LPPMs are more effective than traditional hardwired models when the testing data cannot be fully characterized a-priori by the training data.

To summarize, our contributions are:
\smallskip

\noindent\textbf{$\bullet$} We empirically show that hardwiring the characteristics of a dataset into Location Privacy Preserving Mechanisms~\cite{Quant2011SP,Opt2012CCS,herrmann2013optimal, OptGeoInd2014CCS,Elastic2015PETS,Semantics2016PETS,shokri2017privacy,chatzikokolakis2017efficient,oya2017back,theodorakopoulos2014prolonging} yields mechanisms that do not adequately protect users whose behavior deviates from that observed in training.

\noindent\textbf{$\bullet$} We propose \emph{blank-slate models} for user mobility in location privacy. Contrary to hardwired models, these models treat the user mobility as an \emph{unknown variable} that is learned a-posteriori as the user queries the LBS. Therefore, they enable the design of LPPMs that are effective when the user behavior changes with respect to the one observed when designing the mechanism.

\noindent\textbf{$\bullet$} We leverage a blank-slate sporadic mobility model to develop a new LPPM design technique, that we call Profile Estimation-Based (PEB). PEB-LPPMs adapt to the user behavior by performing a Maximum Likelihood Estimation (MLE) of the mobility profile given past observations, and are suitable for both sporadic and non-sporadic location protection.

\noindent\textbf{$\bullet$} We compare PEB-LPPMs with optimal state-of-the-art designs developed using hardwired sporadic and Markov models. PEB-LPPMs always outperform optimal sporadic hardwired LPPMs, and sometimes they even outperform optimal LPPMs based on Markov models if the training data does not correctly capture the mobility behavior of the users of the testing set.

\noindent\textbf{$\bullet$} To carry out this comparison we extend efficient remapping techniques used in optimal sporadic LPPMs~\cite{chatzikokolakis2017efficient} to build optimal non-sporadic Markov-based LPPMs~\cite{theodorakopoulos2014prolonging,shokri2017privacy}. This considerably reduces the computational cost of building non-sporadic LPPMs and allows us to evaluate them empirically.

The paper is structured as follows. In Sect.~\ref{sec:preliminaries} we introduce our system model and notation, as well as the evaluation framework that we use in the paper. Section~\ref{sec:mobilitymodel} presents the sporadic and Markov mobility models. Then, in Sect.~\ref{sec:hardwired_design}, we explain how previous works use these mobility models, hardwired on training data, to build optimal LPPMs. We train and evaluate these optimal LPPMs with real data in Sect.~\ref{sec:eval1}, showing that there is a gap between their theoretical performance and their actual performance in the testing set. We introduce blank-slate models and our technique to develop PEB-LPPMs in Sect.~\ref{sec:blankslate}, and evaluate it in Section~\ref{sec:eval2}. Finally, Sect.~\ref{sec:relwork} summarizes related work and Sect.~\ref{sec:conclusions} concludes.

\section{Overview of the Location Privacy Problem}
\label{sec:preliminaries}

In this section, we first provide an abstraction of the location privacy problem and introduce our notation. Then, we present our framework for design and evaluation of LPPMs. 

\subsection{Problem Statement and Notation}
\label{sec:notation}

As in previous works~\cite{Quant2011SP,chatzikokolakis2017efficient,Opt2012CCS,oya2017back}, we consider the scenario where an individual, the \emph{user}, sends queries to an LBS provider and receives responses with the information she desires. We assume that there is a passive \emph{adversary} observing the locations inside the user queries, who can be either an honest-but-curious LBS or an eavesdropper. The adversary's goal is to infer private information from the locations in user queries~\cite{Krumm07,GambsKC11}. To protect herself, the user obfuscates her locations using an LPPM, and sends these fake locations in the queries. By doing so, the user trades in quality of service for privacy.

\begin{figure} 
      \centering
      \begin{minipage}[b]{0.9\linewidth}
	\includegraphics[width=\columnwidth]{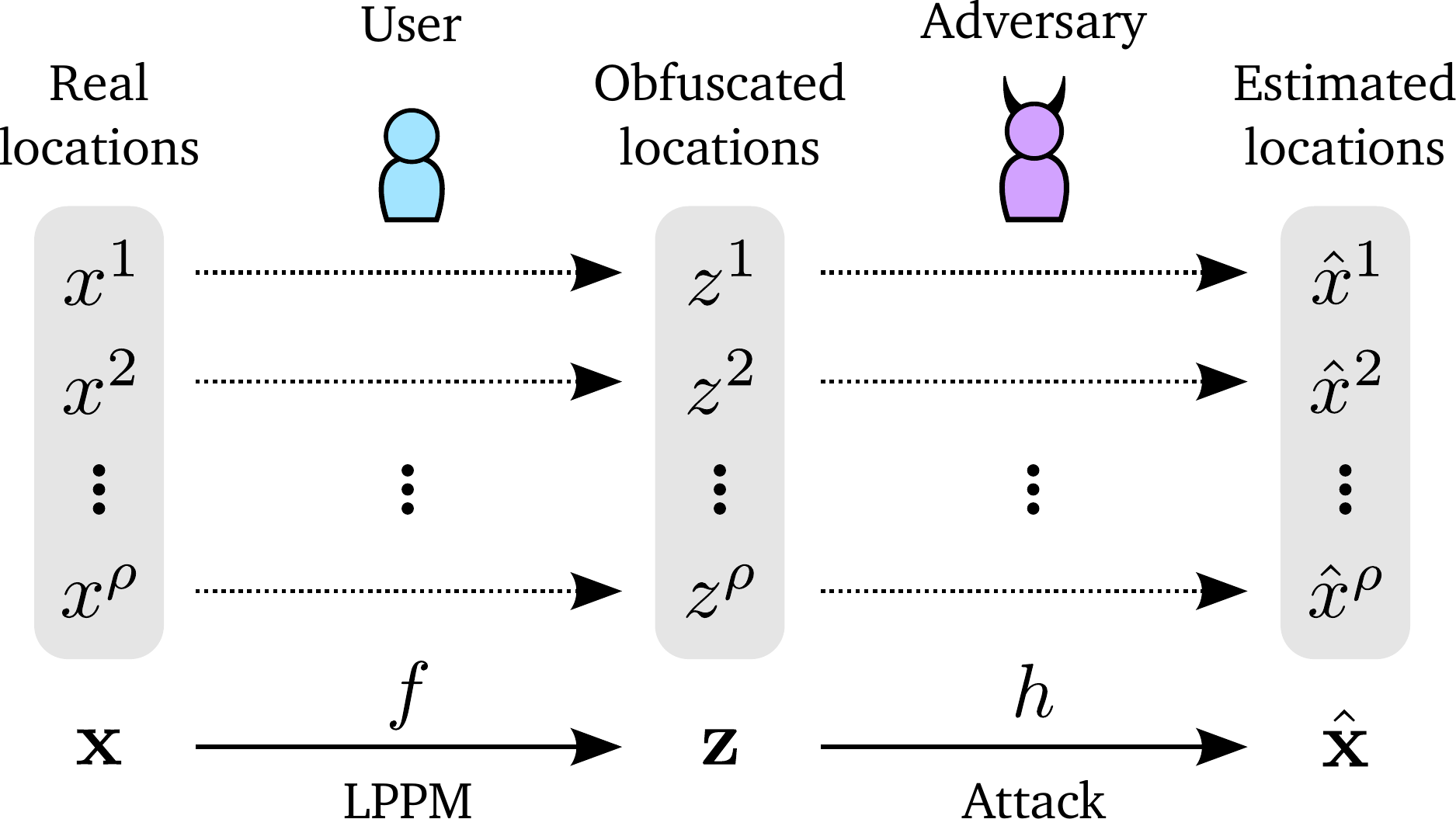}
	\caption{Abstraction of the location privacy problem.}
	\label{fig:model_general}
      \end{minipage}
\end{figure}

We illustrate the location privacy problem in Fig.~\ref{fig:model_general}. We use $\rho$ to denote the total number of queries sent by the user to the LBS, and refer to each query by its query number $r\in\{1,2,\dots,\rho\}$. We use $x^r\in\mathcal{X}$ to denote the real location associated with the $r$-th query, i.e., the location the user wants to query about. We use $\mathbf{x}\doteq[x^1,\dots,x^\rho]\in\mathcal{X}^\rho$ to denote the vector of all the real locations, and $\mathbf{x}^r\doteq[x^1,\dots,x^r]\in\mathcal{X}^r$ to denote the vector of all the real locations up to query number $r$. Likewise, we use $z^r\in\mathcal{Z}$ to denote the $r$-th fake location reported and define the vectors $\mathbf{z}$ and $\mathbf{z}^r$. The real and fake locations are also called input and output locations respectively. Finally, we use $\hat{x}^r\in \hat{\mathcal{X}}$ to denote the adversary's estimation of $x^r$.

In this work, we assume that $\mathcal{X}$, $\mathcal{Z}$ and $\hat{\mathcal{X}}$ are discrete sets of locations (i.e., the users can only report locations in a grid). We do this for computational simplicity and for compatibility with previous proposals \cite{shokri2017privacy, Opt2012CCS,theodorakopoulos2014prolonging}. However, all of our findings can be extended to other scenarios (e.g., $\mathcal{Z}=\mathbb{R}^2$ is the plane~\cite{oya2017back}, $\mathcal{Z}$ is a discrete set of cloaking regions~\cite{gruteser2003anonymous}, or a powerset of points of interest~\cite{herrmann2013optimal}). We use $x_1,x_2,\dots,x_{|\mathcal{X}|}$ to denote each of the discrete locations in $\mathcal{X}$. Finally, we use $p$ generally to denote the probability mass function of a discrete random variable, or the probability density function when the variable is continuous. $\Exp{\cdot}$ denotes the expectation.

Now, we explain how real, obfuscated, and estimated locations are generated. The real locations $\mathbf{x}$ are chosen by the user as she queries the LBS. 
In some scenarios, the user makes a \emph{sporadic} usage of the LBS (e.g., location check-in, location-tagging, or applications for finding nearby points-of-interest or friends). This means that the real locations of two queries (e.g., $x^r$ and $x^s$, with $r\neq s$) are not temporally dependent. In other scenarios, however, the location of the user in consecutive check-ins is correlated (e.g., a user that reports her location frequently, such as running apps or WhatsApp's live location sharing). 

In order to generate obfuscated locations $\mathbf{z}$ from the real locations $\mathbf{x}$ the user employs an LPPM $f$. We study the \emph{online} location privacy setting, in which the user expects to get the service from the LBS right away. In this case, the LPPM is modeled as a probabilistic function that maps a real location $x^r\in\mathcal{X}$, and possibly other information available to the user up to that point (i.e., $\mathbf{x}^{r-1}$ and $\mathbf{z}^{r-1}$), to a value $z^r\in\mathcal{Z}$. We use $f$ to denote the probability density function that characterizes the LPPM. Hence, we can write $p(\mathbf{z}|\mathbf{x})$ as
\begin{equation} \label{eq:online}
 p(\mathbf{z}|\mathbf{x})=\prod_{r=1}^\rho p(z^r|\mathbf{z}^{r-1},\mathbf{x})=\prod_{r=1}^\rho f(z^r|\mathbf{z}^{r-1},\mathbf{x}^{r})\,.
\end{equation}
where the first equality is the chain rule of probability and the second equality reflects the online setting assumption, i.e., the user generates $z^r$ given $\mathbf{x}^r$ and $\mathbf{z}^{r-1}$, but independently of future locations $x^{r+1}$, $x^{r+2}$, etc. We also refer to $f$ as the \emph{obfuscation mechanism}.

Finally, the adversary generates the estimated locations using an attack $h$. We assume that the adversary knows the obfuscation mechanism $f$ and she uses it to design her attack $h$. We treat $h$ as a deterministic function that takes a vector of obfuscated locations $\mathbf{z}^r$ and produces an estimate $\hat{x}^s$ of a (possibly past) real location $x^s$ ($s\leq r$).  We use $\hat{x}^s(\mathbf{z}^r)$ to denote the estimate produced from $\mathbf{z}^r$ using $h$.  We do not consider randomized attacks, since the goal of the adversary is to choose her estimation so as to minimize a specific privacy metric, which can be achieved with deterministic attacks.

\vspace{3pt}\noindent\textbf{LPPM types:} Depending on how much information they use to generate obfuscated locations, LPPMs can offer stronger privacy guarantees at the cost of introducing complexity in the design. In this paper we study the following LPPM types that can accommodate all previous proposals in the literature:
\begin{enumerate}
 \item \emph{Full LPPMs} are the most generic LPPM in the online location privacy setting (see \eqref{eq:online}), i.e., $f(z^r|\mathbf{z}^{r-1},\mathbf{x}^r)$. They generate each obfuscation location $z^r$ (perhaps randomly) using all the information available to the user, i.e., the previous and current input locations $\mathbf{x}^r$, and the previously released obfuscated locations $\mathbf{z}^{r-1}$. 
 \item \emph{Output-based LPPMs}, $f(z^r|\mathbf{z}^{r-1},x^r)$, generate the obfuscated location using only the current real location $x^r$ and all the previous obfuscated locations $\mathbf{z}^{r-1}$. These are a sub-type of full LPPMs.
 \item \emph{Memoryless LPPMs}, $f(z^r|x^r)$, generate each obfuscated location using the current real location $x^r$ only. These are a sub-type of output-based LPPMs.
\end{enumerate}
We note that the framework in~\cite{Quant2011SP} considers LPPMs of the full type in its theoretical setup, but the evaluation studies only memoryless LPPMs. Memoryless LPPMs are used in sporadic location privacy and works that consider a single location release~\cite{herrmann2013optimal,Opt2012CCS,Shokri15,oya2017back,Elastic2015PETS,chatzikokolakis2017efficient}.
Output-based LPPMs are typically used in non-sporadic location privacy works~\cite{theodorakopoulos2014prolonging,shokri2017privacy} and, to the best of our knowledge, no optimal full-LPPM has been proposed due to the computational complexity inherent to its design.

The notation used in the paper is summarized in Table~\ref{tab:notation}.

\begin{table}
\begin{center}
\caption{Summary of notation}
\label{tab:notation}
\begin{tabular}{p{1.7 cm}l}
  \textbf{Symbol} & \multicolumn{1}{l}{\textbf{Meaning}}   \\
  \toprule
  $\rho$ & Total number of queries. \\
  $x^r$ & Real location of the user in the $r$-th query.\\
  $z^r$ & Obfuscated location of the user in the $r$-th query. \\
  $\mathbf{x}^r$ (or $\mathbf{z}^r$) & Vector of real (or obfuscated) locations up to query $r$. \\
  $\mathbf{x}$ (or $\mathbf{z}$) & Vector of all real (or obfuscated) locations. \\
  $\mathcal{X}$ (or $\mathcal{Z}$) & Set of all possible real (or obfuscated) locations. \\
  $h$ & Adversary's attack.\\
  $\hat{x}^s(\mathbf{z}^r)$ & Adversary's estimate of the real location $x^s$ using $\mathbf{z}^r$.\vspace{0.2cm}\\
  
  $f$ & LPPM or obfuscation mechanism (pdf that generates $z^r$).\\
  $f(z^r|\mathbf{z}^{r-1},\mathbf{x}^r)$ & Full LPPM. \\
  $f(z^r|\mathbf{z}^{r-1},x^r)$ & Output-based LPPM. \\
  $f(z^r|x^r)$ & Memoryless LPPM. \vspace{0.2cm}\\
  
  
  $\dQ{x^r,z^r}$ & Quality loss when reporting $z^r$ given $x^r$.\\
  $\Qavg(f,s)$ & Average quality loss metric at query number $r$ \eqref{eq:Qavg}.\\
  $\dP{x^r,\hat{x}^r}$ & Adv.~error when the adversary estimates $x^r$ as $\hat{x}^r$. \\
  $\PAE(f,h,r,s)$ & Avg.~adv.~error of $\hat{x}^s$ given $\mathbf{z}^r$ and attack $h$ \eqref{eq:PAE}. \\
 \end{tabular}
\end{center}
\end{table}

\subsection{Design and Evaluation Framework}
\label{sec:framework}

\begin{figure} 
      \centering
      \begin{minipage}[b]{\linewidth}
	\includegraphics[width=\columnwidth]{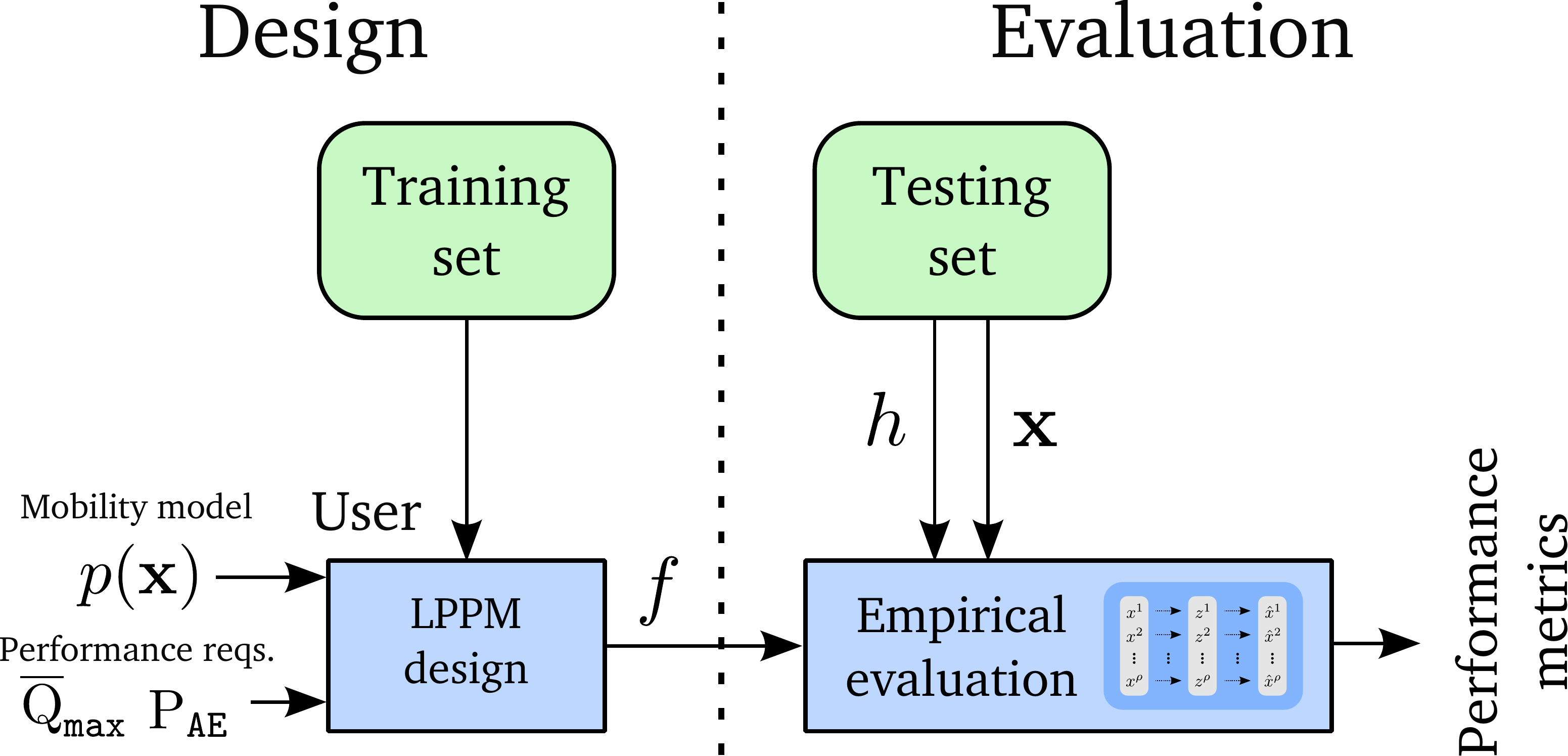}
	\caption{LPPM design and evaluation framework.}
	\label{fig:framework}
      \end{minipage}
\end{figure}

We now describe a framework that instantiates the abstraction above. This framework extends ideas from~\cite{Quant2011SP,chatzikokolakis2017efficient}. It consists of two steps: the design step, where the user designs the LPPM $f$; and the evaluation step, where the performance of $f$ is evaluated empirically. The framework is represented in Fig.~\ref{fig:framework}.

\para{Design Step:}
In this step, the user studies the location privacy problem and builds the LPPM $f$. We assume that the user has access to a \emph{training set}. She derives her design according to some performance requirements, in terms of privacy and utility metrics (e.g., maximizing privacy while keeping the utility level above some bounds). Also, the user does not know the adversary's attack $h$, so she designs the LPPM considering a worst case adversary. In order to compute the privacy and utility metrics, the user needs a model for the joint distribution $p(\mathbf{x},\mathbf{z})=p(\mathbf{x})\cdot p(\mathbf{z}|\mathbf{x})$. The first term, $p(\mathbf{x})$, is the joint distribution of the real locations of the user. The user derives this distribution by training her mobility model with the training set information. The second term, $p(\mathbf{z}|\mathbf{x})$, is determined by the LPPM $f$, as in \eqref{eq:online}.

\para{Evaluation Step:}
In this step, the performance of $f$ against one or more attacks $h$ is assessed empirically using a \emph{testing set}. Following Kerckhoffs principle, we assume that the adversary knows the user LPPM, and uses an optimal attack, i.e., an attack that minimizes the privacy metric. To develop the worst-case attack we assume that the adversary knows mobility statistics about the testing set (e.g., the actual probability distribution that models $\mathbf{x}$), a common assumption in related works~\cite{Quant2011SP,Opt2012CCS,shokri2017privacy,oya2017back}.

The testing set contains real traces of locations $\mathbf{x}$ from a location privacy dataset. The outputs $\mathbf{z}$ are probabilistically generated using $f$ and $\mathbf{x}$. Then, the estimations $\hat{x}^s$ ($s=1,2,\dots$) are calculated using $h$ and $\mathbf{z}$. The privacy and utility performance of the LPPM is assessed empirically based on $\mathbf{x}$, $\mathbf{z}$ and $\hat{x}^s$.

Note that there is a fundamental difference between the design and the evaluation steps, regarding the treatment of the real locations $\mathbf{x}$. The design step is carried out by studying the problem \emph{analytically}, and this is done by assuming a particular mobility model for the real locations $p(\mathbf{x})$. The evaluation step, on the contrary, is carried out \emph{empirically} with real samples of $\mathbf{x}$. Ideally, the user wants her mobility model to closely resemble her real behavior, so that her theoretical analyses translate well into practice. However, finding a realistic model for $\mathbf{x}$ is a very complicated task due to the unpredictability and complexity of user behavior. Notice that this is not an issue for the generation of $\mathbf{z}$. This is because these samples are generated by $p(\mathbf{z}|\mathbf{x})$, which is completely characterized by the obfuscation mechanism and is the same in the design and evaluation steps.

\para{Main Differences with Previous Work:}
This framework takes ideas from the literature, but also adds some contributions. The framework by Shokri et al.~\cite{Quant2011SP} considers an adversary that designs her attack based on the evaluation data, but does not evaluate LPPMs designed to maximize privacy. The separation between training and testing data is considered for the first time in~\cite{chatzikokolakis2017efficient}, but there is no quantification of the privacy loss associated to users' whose mobility profiles diverge from the training data.

In this work, we integrate the training/testing separation as part of the framework. We also consider the selection of a model for the real locations $p(\mathbf{x})$ as a crucial part of the designing step, which was considered as given by previous works~\cite{Quant2011SP,Opt2012CCS,herrmann2013optimal, OptGeoInd2014CCS,Elastic2015PETS,Semantics2016PETS,shokri2017privacy,chatzikokolakis2017efficient,oya2017back,theodorakopoulos2014prolonging}. Finding a suitable theoretical model for the user mobility $p(\mathbf{x})$ and fitting it to the training data is part of the LPPM design process. However, we cannot take for granted that the actual locations of the user in practice $\mathbf{x}$ will follow the theoretical model that she considered for design, and thus the performance of the LPPM in practice might differ from the theoretical performance.

\subsection{Performance Metrics}

We quantify the performance of LPPMs using privacy and utility (or quality loss) metrics. In this work, we use the average quality loss as utility metric, and the average adversary error as privacy metric. These metrics are the most popular in the user-centric location privacy literature~\cite{Opt2012CCS,GeoInd2013CCS,OptGeoInd2014CCS,Elastic2015PETS,chatzikokolakis2017efficient,oya2017back,Semantics2016PETS,Quant2011SP,theodorakopoulos2014prolonging,shokri2017privacy}. We now define these metrics, and explain how to compute them analytically given a model of $p(\mathbf{x})$, and empirically given samples of pairs $(\mathbf{x},\mathbf{z})$.

\para{Utility Metric: Average Quality Loss.} 
The average quality loss measures how much quality the user loses on average by reporting obfuscated locations instead of real ones~\cite{Opt2012CCS,GeoInd2013CCS,OptGeoInd2014CCS,Elastic2015PETS,chatzikokolakis2017efficient,oya2017back,shokri2017privacy,theodorakopoulos2014prolonging}. Let $d_Q(x,z)$ be a point-to-point \emph{distance function} that measures the loss incurred by revealing $z$ when the real location is $x$. The average loss at query $r$ given LPPM $f$ is
\begin{equation} \label{eq:Qavg}
 \Qavg(f,r)\doteq\Exp{d_Q(x^r,z^r)}\,,
\end{equation}
where the expectation is taken over realizations of $x^r$ and $z^r$. Given a distribution $p(\mathbf{x})$, we can compute this metric theoretically as
\begin{equation} \label{eq:Qavg_theo}
 \Qavg^\texttt{theo}(f,r)= \sum_{x^r\in\mathcal{X}} \sum_{z^r\in\mathcal{Z}} p(x^r) \cdot p(z^r|x^r) \cdot d_Q(x^r,z^r)\,,
\end{equation}
where $p(x^r)$ and $p(z^r|x^r)$ can be obtained analytically from $p(\mathbf{x})$ and $p(\mathbf{z}|\mathbf{x})$.

Empirically, we can compute this metric by averaging the distance between $x^r$ and $z^r$ over multiple simulations, i.e.,
\begin{equation}
 \Qavg^\texttt{prac}(f,r)=\Expemp{d_Q(x^r,z^r)}\,,
\end{equation}
where $\Expemp{\cdot}$ denotes the empirical mean.

The typical choice for the distance function $d_Q(\cdot)$ is the Euclidean distance. However, $d_Q(\cdot)$ can be tailored to the particular application where we want to provide location privacy. For example, in an application to find nearby points of interest within a city, the Manhattan distance is appropriate to measure the walking distance to go from $x$ to $z$. In that case, $\Qavg$ would represent the average amount of extra meters that the user has to walk to reach the desired point of interest. In a ride-sharing app, however, $d_Q(x,z)$ can represent the extra time or money that the user loses by reporting $z$ instead of her real location $x$. We can also use semantic metrics based on the location tags of $x$ and $z$, etc.

\para{Privacy Metric: Average Adversary Error.} 
The average adversary error is defined as the mean error incurred by an adversary that estimates the user real locations using an attack $h$~\cite{Quant2011SP,Opt2012CCS,GeoInd2013CCS,OptGeoInd2014CCS,Elastic2015PETS,chatzikokolakis2017efficient,oya2017back,Semantics2016PETS,shokri2017privacy,theodorakopoulos2014prolonging}. Let $d_P(x,\hat{x})$ be a function that quantifies how much privacy the user has when her real location is $x$ and the location estimated by the adversary is $\hat{x}$. Typically, $d_P(\cdot)$ is the Euclidean distance, but it can adapted to a particular application. Consider that the adversary has observed $r$ outputs ($\mathbf{z}^r$) and wants to estimate the location $x^s$ with $s\leq r$. For this, she uses an attack $h$ that produces an estimation $\hat{x}^s(\mathbf{z}^r)$. The average adversary error at query $r$ regarding $x^s$ can be defined as
\begin{equation} \label{eq:PAE}
 \PAE(f,h,r,s)\doteq\Exp{d_P(x^s,\hat{x}^s(\mathbf{z}^r))}\,,
\end{equation}
where the expectation is taken over $x^s$ and $\mathbf{z}^r$ (the attack is deterministic, i.e., $\hat{x}^s$ is a function of $\mathbf{z}^r$).
Given a mobility model $p(\mathbf{x})$, this metric can be computed analytically as
\begin{equation} \label{eq:PAE_theo}
 \PAE^\texttt{theo}(f,h,r,s)=\sum_{\mathbf{x}^r\in\mathcal{X}^r} \sum_{\mathbf{z}^r\in\mathcal{Z}^r} p(\mathbf{x}^r) p(\mathbf{z}^r|\mathbf{x}^r) d_P(x^s,\hat{x}^s(\mathbf{z}^r)) \,.
\end{equation}

Empirically, for each realization of $\mathbf{x}$ and $\mathbf{z}$, we obtain the adversary estimation $\hat{x}^s(\mathbf{z}^r)$, and then compute the average adversary error as
\begin{equation}
 \PAE^\texttt{prac}(f,h,r,s)=\Expemp{d_P(x^s,\hat{x}^s(\mathbf{z}^r))}\,.
\end{equation}

We acknowledge that there are other privacy metrics, e.g., the conditional entropy~\cite{oya2017back} and geo-indistinguishability~\cite{GeoInd2013CCS}. In our empirical evaluation in Sect.~\ref{sec:discussion} we discuss how our findings affect those metrics.

\section{Mobility Models for LPPM Design}
\label{sec:mobilitymodel}

As we explained before, in order to design LPPMs, the user needs to assume a model that characterizes her mobility behavior, i.e., a model for $p(\mathbf{x})$. In this section, we explain the main mobility models assumed in the literature: the \emph{sporadic} mobility model, and the \emph{Markov} model (non-sporadic).
We do not claim that there is a \emph{correct} mobility model for $p(\mathbf{x})$ that the user should follow. However, it is true that LPPMs optimized for a certain model will perform better when the actual user location traces follow such model. In other words, models that are closer to the real behavior of the user are more \emph{useful}.

\subsection{Sporadic Model}

The sporadic location privacy model assumes that the real locations of the user in two different queries, i.e., $x^r$ and $x^s$, are not temporally dependent. As we argued before, this makes sense in some scenarios where the user requests information from the LBS infrequently (e.g., a user that queries for the weather in her area is not likely to perform the another query in a short period of time). 

The sporadic model characterizes $p(\mathbf{x})$ using a parameter called the \emph{mobility profile}, denoted by $\pi$ (Fig.~\ref{fig:models}, left, and Fig.~\ref{fig:table_models}). The mobility profile is an abstraction that represents the long-term user behavior, i.e., the probability with which the user visits each location $x\in\mathcal{X}$. Thus, given $\pi$, we can write
\begin{equation}
 p(\mathbf{x}|\pi)=\prod_{r=1}^\rho p(x^r|\pi)=\prod_{r=1}^\rho \pi(x^r)\,,
\end{equation}
where we have used $\pi(x)$ to denote the probability that the user's real location is $x$ given the profile $\pi$.

This model has been widely used in the literature~\cite{Opt2012CCS,OptGeoInd2014CCS,oya2017back,chatzikokolakis2017efficient}, mainly for its simplicity: using the fact that two check-ins $x^r$ and $x^s$ are independent allows the user to design LPPMs that only need the current input $x^r$ to generate the next output $z^r$.

\subsection{Continuous Model: Markov}

In some scenarios, the sporadic model for user mobility is not appropriate. For example, when a user queries the LBS continuously (e.g., live location sharing in social networks), we cannot assume that the location $x^{r+1}$ is independent of the previous one $x^r$ (e.g., because physical constraints such as the user speed or roads existence and direction). In those cases, continuous models that specify the dependencies between the real locations are more adequate to design LPPMs.

\begin{figure}
      \centering
      \begin{minipage}[b]{0.45\linewidth}
      \centering
	\includegraphics[height=3cm]{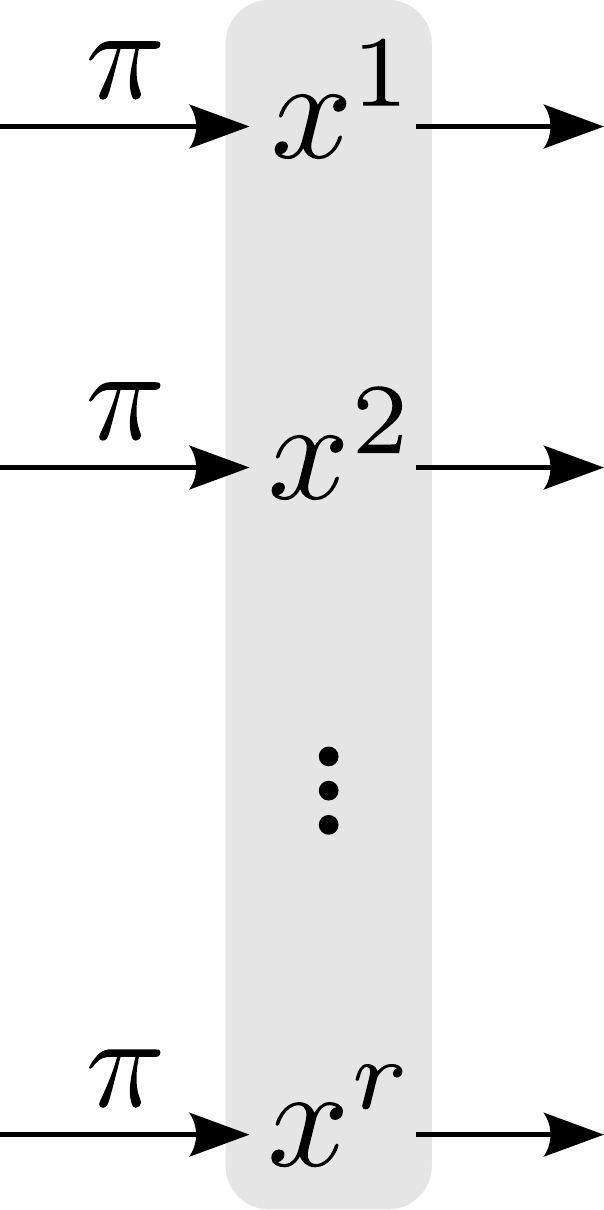} \quad
	\includegraphics[height=3cm]{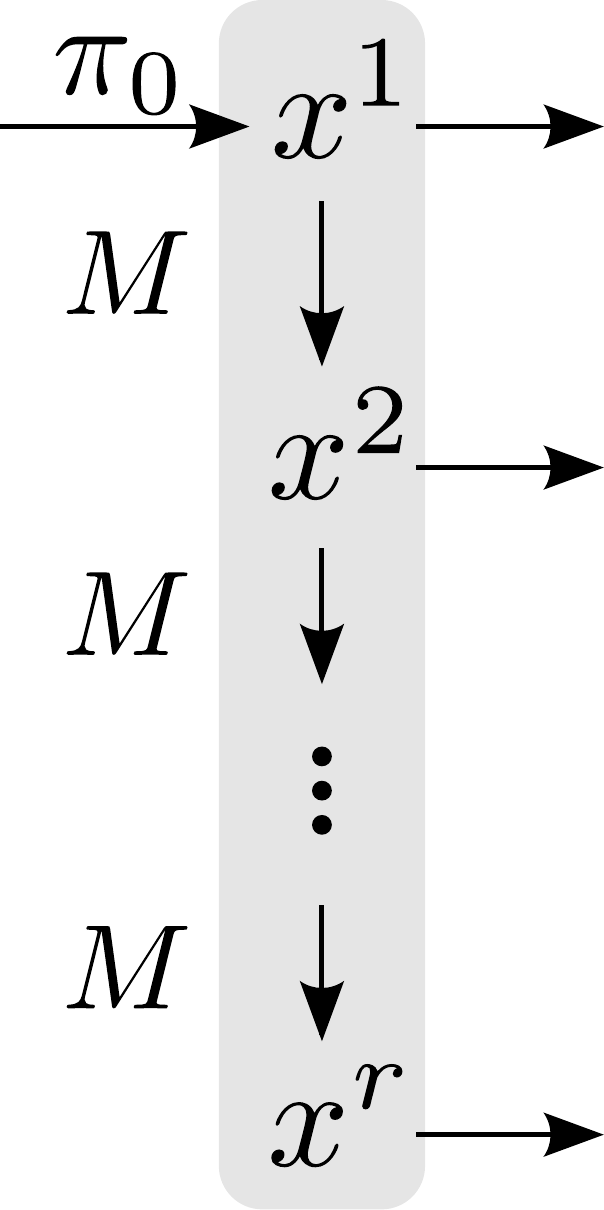}
	\caption{Sporadic (left) and Markov (right) models of user mobility.}
	\label{fig:models}
      \end{minipage}
      \hfill
      \begin{minipage}[b]{0.47\linewidth}
      \centering
	\begin{tabular}{c|c}
	\textbf{} & \textbf{Meaning} \\ \hline
	$\pi$ & Mobility profile \\
	 & $p(x^r|\pi)=\pi(x^r)$ \\ \hline
	$\pi_0$ & Initial mob.~prof. \\
	& $p(x^1|\pi_0)=\pi_0(x^1)$ \\
	$M$ & Transition matrix. \\
	& $p(x^{r+1}|x^r,M)=$ \\
	& $M(x^{r+1}|x^r)$
	\end{tabular}
	\caption{Notation and meaning of the model parameters.}
	\label{fig:table_models}
      \end{minipage}
\end{figure}

The most typical model in this scenario is the Markov model. As its name suggests, this model characterizes $x^r$ as a Markov chain. More specifically, Markov models are defined by two parameters: an initial mobility profile $\pi_0$, and a \emph{transition matrix} $M$ (Fig.~\ref{fig:models}, right, and Fig.~\ref{fig:table_models}). The initial profile models the probability of the first location of the user, i.e., $p(x^1|\pi_0)=\pi_0(x^1)$. The transition matrix $M$ is a $|\mathcal{X}|\times|\mathcal{X}|$ matrix whose $(r,r-1)$-th element characterizes $p(x^r_i|x^{r-1}_j)$, regardless of $r>1$. We use $M(x^r|x^{r-1})$ to denote the probability that the user transitions from location $x^{r-1}$ to $x^r$ according to the matrix $M$. The probability of a trace $p(\mathbf{x})$ according to the Markov model is thus
\begin{align*}
 p(\mathbf{x}|\pi_0,M)&=\prod_{r=1}^\rho p(x^r|\mathbf{x}^{r-1},\pi_0,M) \nonumber \\
		      &=\pi_0(x^1) \cdot \prod_{r=2}^\rho M(x^{r}|x^{r-1})\,.
\end{align*}

The Markov model has been widely used in non-sporadic location privacy, due to its simplicity~\cite{theodorakopoulos2014prolonging,shokri2017privacy}. Note that in the Markov model, the user's mobility behavior only depends on her current location, and not the past trace. It is possible to define more complicated models for continuous location release (e.g., characterize $p(x^r|\mathbf{x}^{r-1})$), but since these are rarely used, we do not consider them in this work.

\subsection{Hardwiring Training Data into the Mobility Model}

After the user chooses one model to design her LPPM, she has to decide how to estimate the parameters of that model (i.e., $\pi$ in the sporadic model, $\pi_0$ and $M$ in the Markov model). In the literature, to the best of our knowledge, all of the proposals rely on some training information to determine these parameters~\cite{Quant2011SP,shokri2011quantifying,Opt2012CCS,OptGeoInd2014CCS,oya2017back,chatzikokolakis2017efficient,shokri2017privacy,theodorakopoulos2014prolonging,Semantics2016PETS}. After this training phase, the model parameters remain fixed during the evaluation. We call the models that are built in this way \emph{hardwired} models. Hardwired models are tailored to the training data \emph{a-priori} during training, and their parameters are never updated or adapted for users that deviate from the training data behavior. Therefore, the LPPMs designed with these models will be optimal in practice if the users' behavior is perfectly captured by the training data. If this is not the case, \emph{or if the training data is insufficient or nonexistent}, it is reasonable that the LPPMs designed with hardwired models will perform worse than expected. We confirm this conjecture later in Section~\ref{sec:eval1}.

\section{LPPM Design in Hardwired Models}
\label{sec:hardwired_design}

In this section, we overview previous approaches to design LPPMs leveraging hardwired models for user mobility. We consider \emph{optimal} LPPM designs, i.e., defense mechanisms that, under a certain mobility model, maximize the privacy metric $\PAE$ against the best possible attack given a constraint on the maximum average loss $\Qavg$ allowed. We note that, given a mobility model, there is a \emph{familiy} of LPPMs that are \emph{all optimal} (i.e., all of them achieve the maximum $\PAE$ given a constraint on $\Qavg$), as proven in~\cite{oya2017back}. However, there are no universally optimal LPPMs \emph{in practice}, i.e., when evaluated with testing data against an optimal attack. Thus, it is important to keep in mind that, even if two members of a family of optimal LPPMs perform equally in theory, they might perform differently in practice.

We explain LPPM design for the $r$-th release: the user is at location $x^r$ and wants to query the LBS by releasing an obfuscated location $z^r$. The user knows all her previous real and obfuscated locations, i.e., $\mathbf{x}^{r-1}$ and $\mathbf{z}^{r-1}$, and the LBS/adversary knows the previously released locations $\mathbf{z}^{r-1}$. The optimal LPPM design problem can be written mathematically as
\begin{equation} \label{eq:optimal_design}
  \begin{aligned}
  f = & \underset{f}{\text{ argmax}} & & \min_h \PAE(f,h,r,r)\,, \\
  & \text{ subject to} & & \Qavg(f,r)\leq \Qmax. \\
  \end{aligned}
\end{equation}
Note that $f$ must satisfy some additional constraints since it is a probability density function, but we have omitted those from \eqref{eq:optimal_design} for simplicity. Also, we have considered just the case where the user wants to protect her current location at time $r$. We note however that the user could set other goals, like trying to protect the privacy of future location releases $\PAE(f,h,r,s)$ for $s>r$, past locations ($s<r$), or a combination of both. Our findings could be adapted to such cases, but we do not study them here for simplicity and space restrictions.

We also limit ourselves to optimal output-based LPPMs, i.e., defenses that can be characterized by $f(z^r|\mathbf{z}^{r-1},x^r)$ and do not depend on previous inputs $\mathbf{x}^{r-1}$. We do this to avoid the computational issues that stem from the fact that, in order to guarantee that an LPPM is optimal, the user has to assess its privacy against an optimal attack. In order to do this with a full-LPPM $f(z^r|\mathbf{z}^{r-1},\mathbf{x}^r)$, she has to characterize the posterior probability of the secret locations after releasing the obfuscated locations, i.e., $p(\mathbf{x}^r|\mathbf{z}^{r-1})$. If $x\in\mathcal{X}$ and $\mathcal{X}$ is discrete, this requires handling $|\mathcal{X}|^r$ values, which quickly becomes unfeasible for any computer (e.g., in a small map with $|\mathcal{X}|=200$ discrete locations, if we represent a float with 4 bytes, to protect only $r=8$ locations we would need over 1 million Terabytes). Since measuring the privacy against an optimal adversary is unfeasible in full-type LPPMs, we do not consider them in our design approaches.

Note that this computational issue is not a problem in output-based LPPMs. This is because, in this case, to assess the performance against an optimal adversary the user internally computes $p(x^r|\mathbf{z}^{r-1})$. She only needs to handle $|\mathcal{X}|$ parameters for this, since $\mathbf{z}^{r-1}$ have been seen in the past by both the user and the adversary, so they can be treated as fixed parameters at time $r$.

Below, we explain how to compute optimal LPPMs in the sporadic and Markov hardwired models.

\subsection{LPPM Design in the Hardwired Sporadic Model}
\label{sec:hardwired_design_sporadic}

In the literature, we find many works that study LPPM design under the sporadic hardwired model for user mobility. Most works consider that the LPPM belongs to the memoryless type $f(z^r|x^r)$, either for tractability~\cite{Opt2012CCS,oya2017back} or because they focus on single queries~\cite{OptGeoInd2014CCS,chatzikokolakis2017efficient}. In Appendix~\ref{sec:proofMemoryless}, we formally prove that, in the hardwired model, a properly designed LPPM of the memoryless type \emph{does not provide less privacy} than an LPPM of the full type $f(z^r|\mathbf{z}^{r-1},\mathbf{x}^r)$. This means that considering full-type or output-based LPPMs just complicates the problem and does not provide any advantage over memoryless LPPMs.

There are two main approaches to compute optimal LPPMs in sporadic models:

\para{Linear Programming Approaches.}
Shokri et al.~provide a technique to design optimal LPPMs given any pair of functions $d_P(\cdot)$ and $d_Q(\cdot)$~\cite{Opt2012CCS}. This approach consists on solving a linear program, which can only be done, for computational reasons, if the spaces of real ($\mathcal{X}$) and obfuscated ($\mathcal{Z}$) locations are discrete. The program receives the mobility profile $\pi$ which determines the distribution of $x^r$, and returns an optimal LPPM $f(z^r|x^r)$. If the number of discrete locations is $N$, the linear program contains $N(N+1)$ bounded variables, $N^2+1$ inequality constraints, and $N$ equality constraints. Therefore, finding an optimal obfuscation mechanism using linear programming is only feasible if the number of discrete locations is modest.

Also, Oya et al.~showed in~\cite{oya2017back} that the algorithm used to solve the linear program greatly affects the performance of the resulting LPPM in terms of other privacy metrics (e.g., the conditional entropy). The recommendation in~\cite{oya2017back} is to use an interior-point algorithm, rather than a simplex algorithm.

\para{Remapping Techniques.}
In~\cite{chatzikokolakis2017efficient}, Chatzikokolakis et al.~propose a technique called \emph{optimal remapping} that provides an average loss improvement for any memoryless LPPM, without reducing privacy. They proposed this method under the hardwired sporadic mobility model. We describe this technique briefly, since we use it in this work. Let $\tilde{f}$ be an obfuscation mechanism, and let $\tilde{z}^r$ be an obfuscated location generated from $x^r$ using such LPPM. Before reporting $\tilde{z}^r$, the user can compute the posterior $p(x^r|\tilde{z}^r)$ using $\pi(x^r)$ and $\tilde{f}$. With this posterior, she can compute an alternative obfuscated location $z^r$:
\begin{equation} \label{eq:remap}
 z^r = \underset{z^r}{\text{argmin }} \sum_{x^r\in\mathcal{X}} p(x^r|\tilde{z}^r) \cdot d_Q(x^r,z^r)\,.
\end{equation}
By reporting $z^r$ (instead of $\tilde{z}^r$), the user achieves a reduction on her average loss (if the mobility profile $\pi$ of the sporadic model used to compute \eqref{eq:remap} is close to her real behavior). Also, note that no information about the previous or current input is used in the remapping (since the posterior is computed only using the current output and $\pi$, which are known to the adversary). This means that, by performing this ``remapping'' from $\tilde{z}^r$ to $z^r$, the privacy of the resulting LPPM cannot decrease.

Later, in~\cite{oya2017back}, Oya et al.~proved that if the distance functions used to measure privacy and utility are the same (i.e., $d_P(\cdot)\equiv d_Q(\cdot)$), the LPPM that results from remapping \emph{any} LPPM is optimal in the hardwired sporadic mobility model. This technique can even be applied to design LPPMs when their output space is the plane $\mathcal{Z}\equiv \mathbb{R}^2$. Overall, solving \eqref{eq:remap} is much faster than solving the linear program mentioned above, although it only yields optimal LPPMs if $d_P(\cdot)\equiv d_Q(\cdot)$.

\subsection{LPPM Design in the Hardwired Markov Model}
\label{sec:hardwired_design_markov}

In the Markov model, the input locations $x^1, x^2, \dots$ are correlated. This creates dependencies between past released locations $\mathbf{z}^{r-1}$ and the current location $x^r$, that the user must take into account when designing the LPPM. 

To the best of our knowledge, the only approach to compute optimal LPPMs under the Markov mobility model consists on solving a linear program~\cite{theodorakopoulos2014prolonging,shokri2017privacy}. We explain this approach, and then extend the remapping techniques of sporadic models so that we can efficiently design optimal LPPMs under the Markov model.

\para{Linear Programming Approaches.}
Theodorakopoulos et al.~\cite{theodorakopoulos2014prolonging} extend the linear programming approach of~\cite{Opt2012CCS} to the non-sporadic location privacy case. They propose a framework where the user can specify which obfuscated location(s) she wants to generate at time $r$, which real locations she wants to protect, and which obfuscated locations were released to the LBS in the past. In their implementation, they specifically consider a Markov model for user mobility. In the case we are studying, where the user wants to release $z^r$ to protect $x^r$ and $\mathbf{z}^{r-1}$ have already been released, the approach works as follows.

For the first release ($r=1$), the user just takes the initial profile $\pi_0(x^1)$ and solves a linear program analogous to the sporadic location privacy one~\cite{Opt2012CCS}. This produces an LPPM $f(z^1|x^1)$ that maximizes the privacy metric given a quality loss constraint. Then, she computes the posterior $p(x^1|z^1)$ using $\pi_0(x^1)$ and Bayes' formula, and uses it to obtain the probability distribution of the next real location given the released location: $p(x^2|z^1)=\sum_{x^1\in\mathcal{X}} M(x^2|x^1) \cdot p(x^1|z^1)$.

For the next releases ($r>1$), the steps are analogous, but they use $p(x^r|\mathbf{z}^{r-1})$ instead of $\pi_0$. Particularly, before the $r$-th query the user knows $p(x^r|\mathbf{z}^{r-1})$. With this probability distribution, the user can solve a linear program to find an optimal LPPM $f(z^r|\mathbf{z}^{r-1},x^r)$. Then, she can compute the posterior using Bayes' formula:
\begin{equation} \label{eq:Markov_posterior}
 p(x^r|\mathbf{z}^r)=\frac{f(z^r|\mathbf{z}^{r-1},x^r)\cdot p(x^r|\mathbf{z}^{r-1})}{\sum_{\tilde{x}^r\in\mathcal{X}} f(z^r|\mathbf{z}^{r-1},\tilde{x}^r)\cdot p(\tilde{x}^r|\mathbf{z}^{r-1})}\,,
\end{equation}
and update it for the next step using the Markov transition matrix:
\begin{equation} \label{eq:Markov_update}
 p(x^{r+1}|\mathbf{z}^r)=\sum_{x^r\in\mathcal{X}}  M(x^{r+1}|x^r) \cdot p(x^r|\mathbf{z}^r)\,.
\end{equation}

In~\cite{theodorakopoulos2014prolonging,shokri2017privacy}, the authors evaluate their LPPMs theoretically, i.e., they compute the average adversary error and average loss that the user would have if she followed the Markov model using the analytical expressions \eqref{eq:Qavg_theo} and \eqref{eq:PAE_theo}. For example, they compute $f(z^2|z^1,x^2)$ for all possible values of $z^2$, $z^1$, $x^2$. Therefore, for computational reasons, they do not evaluate the performance of these LPPMs for more than $r=3$ consecutive locations. During an empirical evaluation, however, one does not need to store all possible values of these variables. Since the past obfuscated locations $\mathbf{z}^{r-1}$ are known both to the user and the adversary, the user can just compute $f(z^r|\mathbf{z}^{r-1},x^r)$ by assuming that $\mathbf{z}^{r-1}$ is fixed. Therefore, the computational cost of computing this Markov-based LPPM in each query is the same as solving the linear program in the sporadic case.

\para{Remapping Techniques.}
Even though the complexity of the linear programming approach in the Markov scenario is the same as in the sporadic scenario, if the number of discrete locations we consider is not small, finding an optimal LPPM is still computationally expensive. To solve this issue, we extend the remapping techniques to the Markov scenario. To the best of our knowledge, this is the first time these techniques are extended beyond the sporadic location privacy scenario.

Assume that, at the time of the $r$-th location release, the user has computed $p(x^r|\mathbf{z}^{r-1})$ according to the Markov model. Let $\tilde{f}$ be any memoryless-LPPM $\tilde{f}(\tilde{z}^r|x^r)$. The user uses this LPPM to generate a temporary $\tilde{z}^r$, and then computes the posterior
\begin{equation}
 p(x^r|\tilde{z}^r,\mathbf{z}^{r-1})=\frac{\tilde{f}(\tilde{z}^r|x^r)\cdot p(x^r|\mathbf{z}^{r-1})}{\sum_{\tilde{x}^r\in\mathcal{X}}\tilde{f}(\tilde{z}^r|\tilde{x}^r)\cdot p(\tilde{x}^r|\mathbf{z}^{r-1})}\,.
\end{equation}

With this posterior, she can then compute the final location that she releases
\begin{equation} \label{eq:remap2}
 z^r = \underset{z^r}{\text{argmin }} \sum_{x^r\in\mathcal{X}} p(x^r|\tilde{z}^r,\mathbf{z}^{r-1}) \cdot d_Q(x^r,z^r)\,.
\end{equation}

This process defines a new LPPM $f(z^r|\mathbf{z}^{r-1},x^r)$. At this point, the user can compute $p(x^{r+1}|\mathbf{z}^r)$ for the next release following \eqref{eq:Markov_posterior} and \eqref{eq:Markov_update}. Computing the LPPM by solving \eqref{eq:remap2} is much faster than solving the linear program explained above. Also, the LPPM that results form the remapping can be shown to be optimal in the hardwired Markov model if $d_P(\cdot)\equiv d_Q(\cdot)$ (c.f.~\cite{oya2017back}).

\section{Evaluation: Optimal Hardwired LPPMs}
\label{sec:eval1}

In this section, we evaluate the optimal LPPMs derived for hardwired models that we described in Sect.~\ref{sec:hardwired_design} using the evaluation framework in Section~\ref{sec:framework}. For readability and clarity, we use the term $\HWsporadic$ to denote a generic LPPM that is optimal under the hardwired \textbf{SP}oradic mobility model~\cite{Opt2012CCS,chatzikokolakis2017efficient,oya2017back}. This LPPM can be computed by following any of the techniques explained in Sect.~\ref{sec:hardwired_design_sporadic}. Likewise, we use $\HWmarkov$ to denote an LPPM that is optimal under the hardwired \textbf{M}ar\textbf{K}ov mobility model~\cite{theodorakopoulos2014prolonging,shokri2017privacy} (we can compute it as explained in Sect.~\ref{sec:hardwired_design_markov}). Note that $\HWsporadic$ and $\HWmarkov$ denote \emph{families} of optimal LPPMs (i.e., there are infinite instantiations of them that meet their optimality conditions).

We perform two different experiments: one to evaluate $\HWsporadic$ in the sporadic location release scenario (Experiment SP), and another one to evaluate $\HWmarkov$ in the continuous location release scenario (Experiment MK). For these experiments, we consider three datasets, two different instantiations of $\HWsporadic$ and $\HWmarkov$, and two optimal attacks. We explain these choices below. Table~\ref{tab:terminology} summarizes the new terminology of this evaluation, and Table~\ref{tab:experiments1} shows the configuration of our experiments.

\begin{table}
\begin{center}
\caption{Terminology for the experiments.}
\label{tab:terminology}
\begin{tabular}{r p{6.8cm}}
   $\HWsporadic$		& Family of optimal LPPMs developed with the hardwired \emph{sporadic} mobility mode (Sect.~\ref{sec:hardwired_design_sporadic}). \\
   $\HWmarkov$			& Family of optimal LPPMs developed with the hardwired \emph{Markov} mobility mode (Sect.~\ref{sec:hardwired_design_markov}). \\ \hline
   $\LHsp$			& Optimal location hiding LPPM from the $\HWsporadic$ family.\\
   $\Exposp$			& Optimal exponential LPPM from the $\HWsporadic$ family.\\
   $\LHmk$			& Optimal location hiding LPPM from the $\HWmarkov$ family.\\
   $\Expomk$			& Optimal exponential LPPM from the $\HWmarkov$ family.\\
 \end{tabular}
\end{center}
\end{table}

\begin{table}
\begin{center}
\caption{Summary of the experiments to evaluate hardwired LPPMs.}
\label{tab:experiments1}
\begin{tabular}{c|c|c}
				& \textbf{Experiment SP} 		& \textbf{Experiment MK}\\
   Evaluation target		& $\HWsporadic$				& $\HWmarkov$ \\ \hline
   \multirow{3}{*}{Datasets} 	& Gowalla (shuffled) 			& Gowalla \\
				& Brightkite (shuffled)			& Brightkite \\
    				& 					& TaxiCab \\ \hline			
   Distance function		& \multicolumn{2}{c}{Manhattan ($d_P\equiv d_Q$)} \\ \hline
   \multirow{2}{*}{LPPM}	& Loc.~Hiding ($\LHsp$)			& Loc.~Hiding ($\LHmk$) \\
				& Exponential ($\Exposp$)		& Exponential ($\Expomk$) \\ \hline
   Attack we evaluate		& Optimal Sporadic 			& Optimal Markov \\
 \end{tabular}
\end{center}
\end{table}

\para{Datasets.} We consider three datasets: Brightkite\footnote{\texttt{https://snap.stanford.edu/data/loc-brightkite.html}}, Gowalla\footnote{\texttt{https://snap.stanford.edu/data/loc-gowalla.html}}, and TaxiCab traces from CRAWDAD.\footnote{\texttt{https://crawdad.org/epfl/mobility/20090224/}} Each dataset contains location traces identified by the user ID, latitude, longitude, and timestamp. We take user check-ins inside the San Francisco region between latitude coordinates $37.5500$ and $37.8010$, and longitude coordinates $-122.5153$ and $-122.3789$. Then, we quantize the area into $25\times 10$ cells and consider the centers of these cells as our alphabets $\mathcal{X}=\mathcal{Z}=\hat{\mathcal{X}}$, as in~\cite{theodorakopoulos2014prolonging,shokri2017privacy}. 

Gowalla and Brightkite are examples of datasets with very sparse check-in behavior (e.g., in Gowalla, each user has an average of 60 check-ins during over 20 months of data collection). Thus, in these datasets we separate 20 users that have at least 300 check-ins inside the San Francisco region, regardless of when those check-ins were made, and save the remaining check-ins of all the other users together (around $35\,000$ in Brightkite and $75\,600$ in Gowalla).

\begin{figure} 
  \centering
  \subfloat[Brightkite/Gowalla datasets.]{\includegraphics[height=6cm]{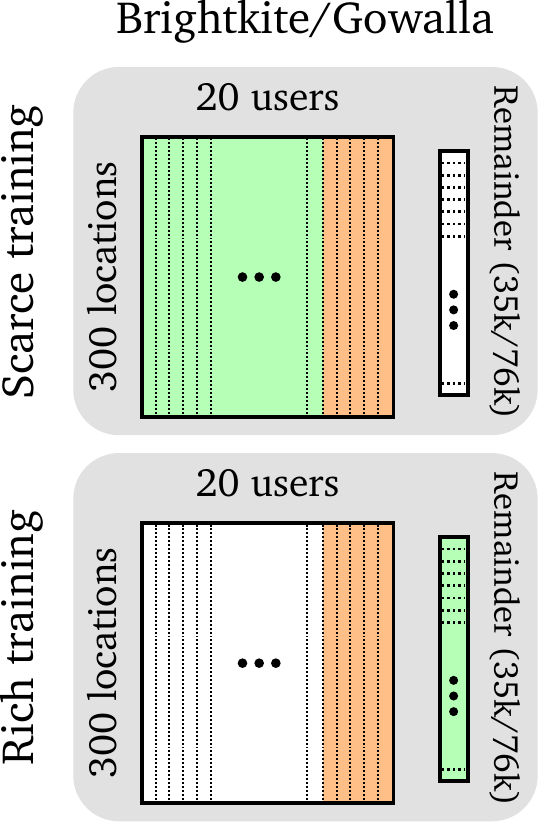}\label{fig:datasets_1}}
  \hfil
  \subfloat[Taxicab dataset.]{\includegraphics[height=6cm]{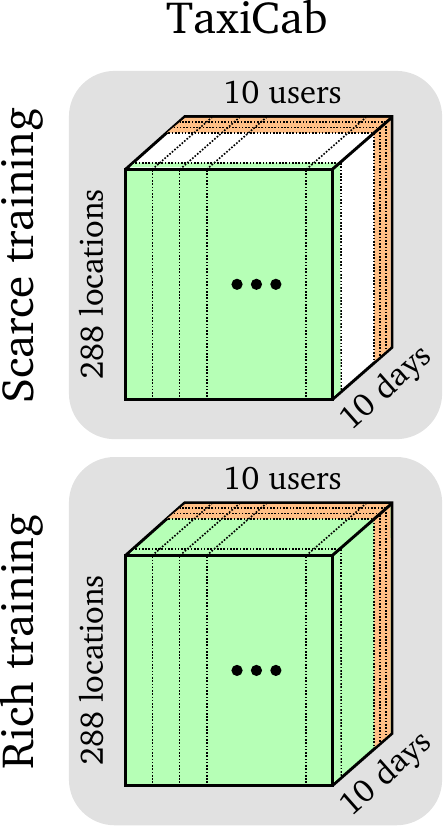}\label{fig:datasets_2}}
  \caption{Processed datasets that we collected. We consider two training settings for each dataset (scarce and rich). For each of these settings, the figure displays the training data in green, and the testing data in orange.}
  \label{fig:datasets}
\end{figure}

Regarding the training/testing separation, we use the last 5 users in these datasets as testing sets, and we consider two training settings: in the first setting, that we call \emph{scarce training}, the users train their LPPMs with the traces of the first 15 users ($4\,500$ locations). In the second setting, that we call \emph{rich training}, each user trains her model using the check-ins of all the other users in the dataset ($35\,000$ in Brightkite, $75\,600$ in Gowalla). This is depicted in Fig.~\ref{fig:datasets_1}.

TaxiCab contains very dense location reports of cabs in the San Francisco region over 30 days. In this case, we organize each user's traces by days, and discard those days where the user remains silent for more than 2 hours. Then, we select 10 users for which we retain at least 10 days. For each trace, we select one check-in for each period of 5 minutes (considering that the user remained in the same location if she did not perform a new check-in in the last 5 minutes). This way, we build, for each user, a set of 10 days with 288 check-ins ($288\cdot 5$ minutes = 1 day). 

In this dataset, we evaluate the performance of each user in her last 3 days. We consider two settings for the training data: in our first setting (scarce), each user uses her first day as training data. In our second setting (rich), each user trains her model using her first 7 days of data (Fig.~\ref{fig:datasets_2}).

\para{Training the LPPMs.} We explain how the users estimate the parameters of the models that they use to build optimal LPPMs. For the LPPMs built using the hardwired sporadic model ($\HWsporadic$), each user computes $\pi$ as a normalized histogram of the training set traces, i.e., she counts the number of check-ins in each location $x\in\mathcal{X}$ in the training data and normalizes by the total number of check-ins.

For LPPMs built using the hardwired Markov model ($\HWmarkov$), each user computes $\pi_0$ as a normalized histogram of the location check-ins in the training data, and builds the transition probabilities $M(x_i|x_j)$ by counting the number of transitions from $x_j$ to $x_i$ in the data, and normalizing. Note that, ideally, the user would compute $\pi_0$ using her first location of each day (in the TaxiCab dataset) but we do not have sufficient data to follow this approach.

\para{LPPM instantiations.} For each family of optimal LPPMs ($\HWsporadic$ and $\HWmarkov$) we test the performance of two different instantiations:
\begin{itemize}
 \item $\LH$ refers to \emph{location hiding}: for each input location $x^r$, the user chooses randomly between revealing her real location $z^r=x^r$ (with probability $\alpha$) or not revealing any information (with probability $1-\alpha$). We model the second case as picking uniformly at random another location of the map. We test 10 values of $\alpha=0, 0.1, 0.2,\dots, 1$ to study the trade-off between $\PAE$ and $\Qavg$. We apply an optimal remapping to this LPPM to make it optimal: the remapping in Sect.~\ref{sec:hardwired_design_sporadic} gives us an $\HWsporadic$ that we denote $\LHsp$, and the remapping in Sect.~\ref{sec:hardwired_design_markov} gives us an $\HWmarkov$ that we denote $\LHmk$.
 \item $\Expo$ is the \emph{exponential} LPPM~\cite{dwork2008differential}: this LPPM reports location $z^r$ with a probability proportional to $\text{exp}(-d_Q(x^r,z^r) \cdot \epsilon)$ (i.e., it has an exponentially decreasing probability of reporting locations that are far from the real location). We test 10 values of $\epsilon=0\text{km}^{-1}, 0.02\text{km}^{-1}, 0.04\text{km}^{-1}, \dots, 0.02\text{km}^{-1}$ to tune the average loss and privacy of this LPPM. We apply an optimal remapping to this LPPM to build an optimal defense (denoted $\Exposp$ and $\Expomk$ for the sporadic and Markov models, respectively).
\end{itemize}

\para{Attacks.} As we mentioned in Section~\ref{sec:framework}, we consider a worst-case adversary which deploys optimal attacks constructed with information about the testing data. The optimal attacks, after observing $z^r$, compute the posterior $p(x^r|\mathbf{z}^r)$ and pick the $\hat{x}^r$ that minimizes the privacy $\PAE$. We consider two attacks: a sporadic-based and a Markov-based attack. These attacks use the actual mobility profiles and transition matrices of the users (i.e., computed from the testing data) to perform their estimation $\hat{x}^r$.

In all of our experiments, we use the Manhattan distance as the distance metric for privacy $d_P(\cdot)$ and utility $d_Q(\cdot)$. We think this is a reasonable choice, since our traces belong to metropolitan areas, where the Manhattan distance between two points is close to the physical distance that a car/person has to traverse to move from one point to the other. We measure distance in kilometers (km), but this could be converted to time (by dividing it by speed) or another metric related to the physical distance between two points. 

We note that, since we chose $d_P(\cdot)\equiv d_Q(\cdot)$, the theoretical performance of any optimal LPPM is $\PAE=\Qavg$, as shown empirically in~\cite{Opt2012CCS} and proven analytically in~\cite{oya2017back}. This means that any optimal LPPM evaluated in the same data used for its training would achieve $\PAE=\Qavg$. This is true for $\LHsp$ and $\Exposp$ against the optimal sporadic attack, and for $\LHmk$ and $\Expomk$ against the optimal Markov attack. We see below that, when these optimal LPPMs are evaluated on a testing set that is different from the training data, they do not achieve this optimal privacy level, i.e., in practice, $\PAE<\Qavg$.

We explain how we generate the plots in our evaluation. Given a particular experiment, user, and LPPM setting, we compute $\Qavg(f,r)$ and $\PAE(f,h,r,r)$ by averaging 40 repetitions of our experiment when users use $\LHsp$ and $\LHmk$, and 20 repetitions when they use $\Exposp$ and $\Expomk$ (i.e., we repeat the process of computing the LPPM, generating obfuscated locations and computing the adversary estimation 20/40 times). Then, we average the performance over $r$ (i.e., $\Qavg\doteq 1/\rho \sum_{r=1}^\rho \Qavg(f,r)$ and $\PAE\doteq 1/\rho \sum_{r=1}^\rho \PAE(f,h,r,r)$). This gives us, for each user that we evaluate, points along their $\PAE$ vs.~$\Qavg$ performance line. Finally, we generate quality loss values linearly spaced between 0 and 4km and, using linear interpolation, compute the average, maximum and minimum privacy over the users for each of those quality loss values. All of our experiments are conducted using Python 3.

\subsection{Experiment SP: Sporadic Hardwired LPPMs}

We evaluate $\LHsp$ and $\Exposp$ against the optimal sporadic-based attack that uses the real mobility profile of the user. We use only Gowalla and Brightkite datasets for this experiment, since Taxicab is more characteristic of non-sporadic mobility behaviors.  For each simulation of this experiment, we randomly shuffle the user traces (i.e., each column in the matrix represented in Fig.~\ref{fig:datasets_1}). We do this to break any possible timing correlation that remains in these datasets and ensure that our evaluation of these LPPMs is fair.

\begin{figure*} 
  \centering
  \subfloat[Brightkite, $\Exposp$.]{\includegraphics[width=0.25\linewidth]{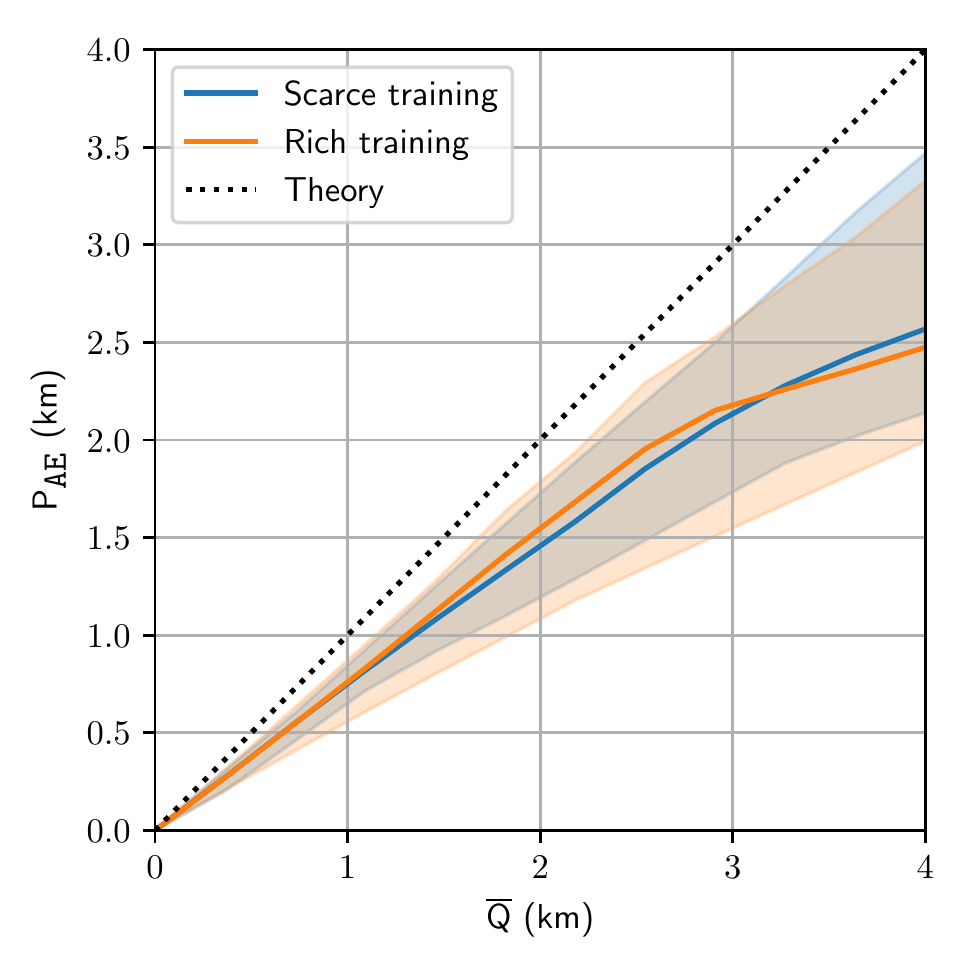}}
  \hfil
  \subfloat[Brightkite, $\LHsp$.]{\includegraphics[width=0.25\linewidth]{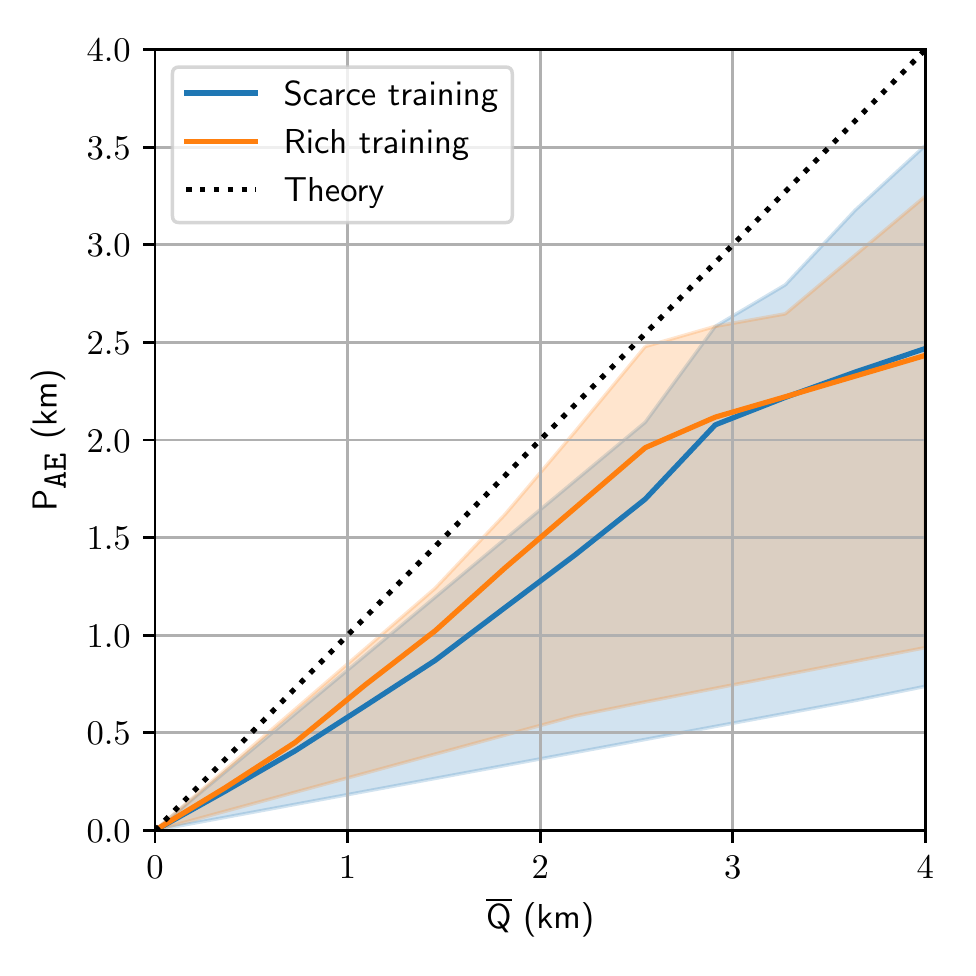}\label{fig:Eval1-Exp1-BKSP}}
  \hfil
  \subfloat[Gowalla, $\Exposp$.]{\includegraphics[width=0.25\linewidth]{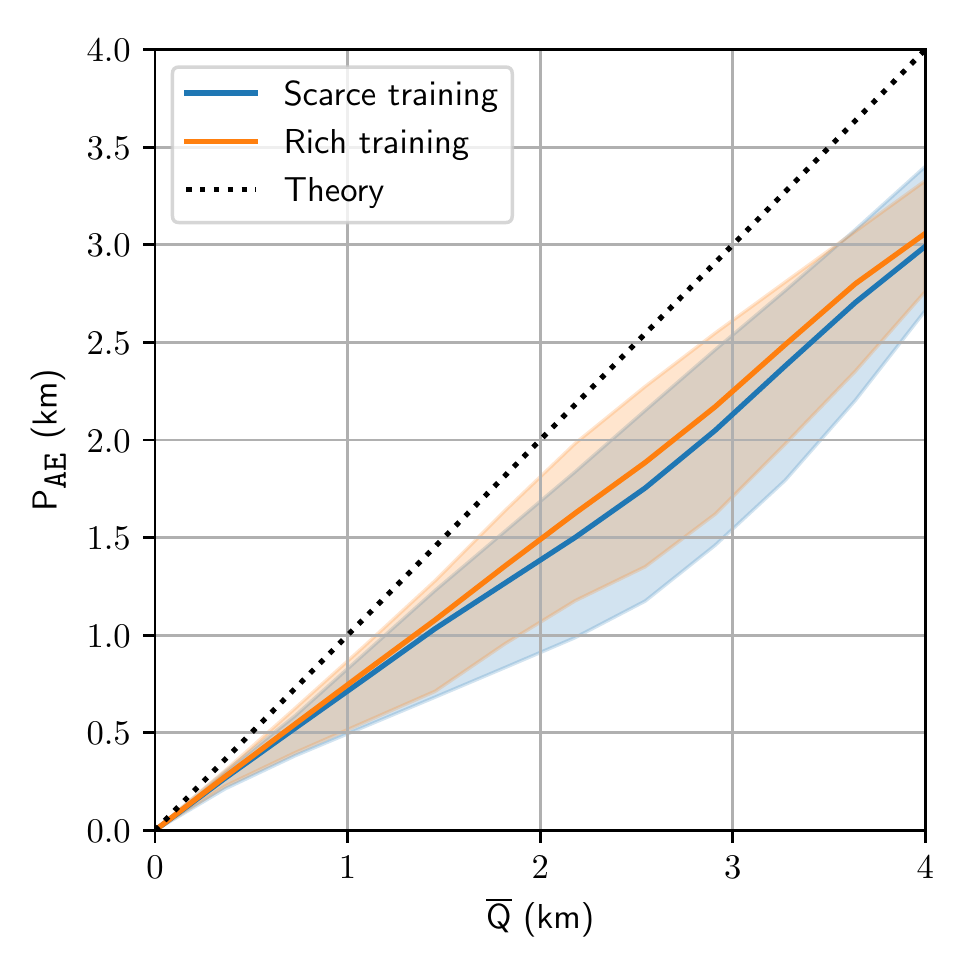}}
  \hfil
  \subfloat[Gowalla, $\LHsp$.]{\includegraphics[width=0.25\linewidth]{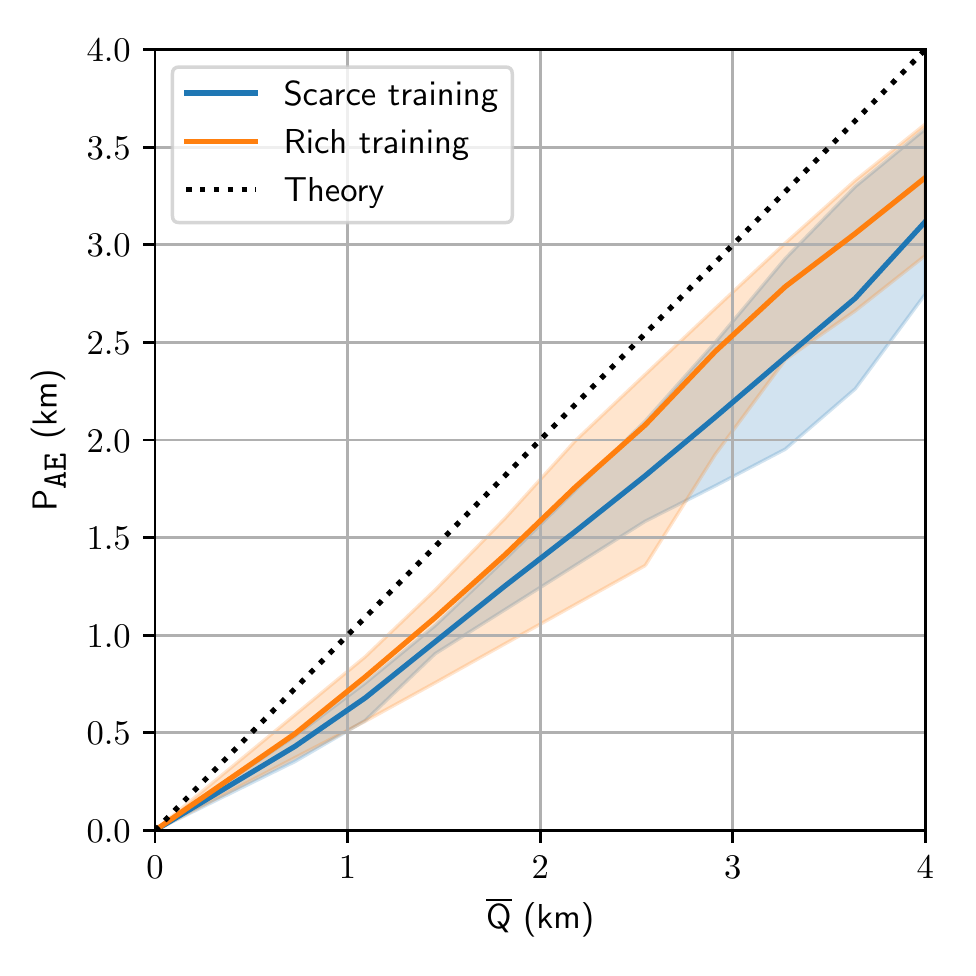}}
  \caption{Experiment SP. Performance of $\Exposp$ and $\LHsp$ against the optimal sporadic attack in Brightkite and Gowalla datasets (with shuffled traces).}
  \label{fig:Eval1-Exp1}
\end{figure*}

Figure~\ref{fig:Eval1-Exp1} shows the results, where the blue and orange lines represent the average privacy of the users when they use the scarce and rich training data, respectively. The shaded area represents the minimum and maximum privacy among the users that we evaluate. We see that, in both datasets, and regardless of the LPPM type, the privacy of the users evaluated with a testing data that differs from the training information is below the theoretical value $\PAE=\Qavg$. Also, training with the rich training set provides more privacy on average, since this dataset has more information about the sporadic check-in behavior of the users ($35\,000-75\,600$ check-ins, versus $4\,500$ check-ins of the scarce dataset). However, this improvement is slight: none of the training sets capture the real user behavior precisely, since both contain data from different users. Some of the users that we evaluate have a behavior that is particularly different from the training data (e.g., lower shaded area in Fig.~\ref{fig:Eval1-Exp1-BKSP}), and thus achieve very low privacy. This experiment shows that training an optimal sporadic LPPM with location data from other users (e.g.,~\cite{chatzikokolakis2017efficient}) is very dangerous from a privacy standpoint.

\subsection{Experiment MK: Markov Hardwired LPPMs}
\label{sec:experimentMK}

\begin{figure*} 
  \centering
  \subfloat[Brightkite, $\Expomk$.]{\includegraphics[width=0.25\linewidth]{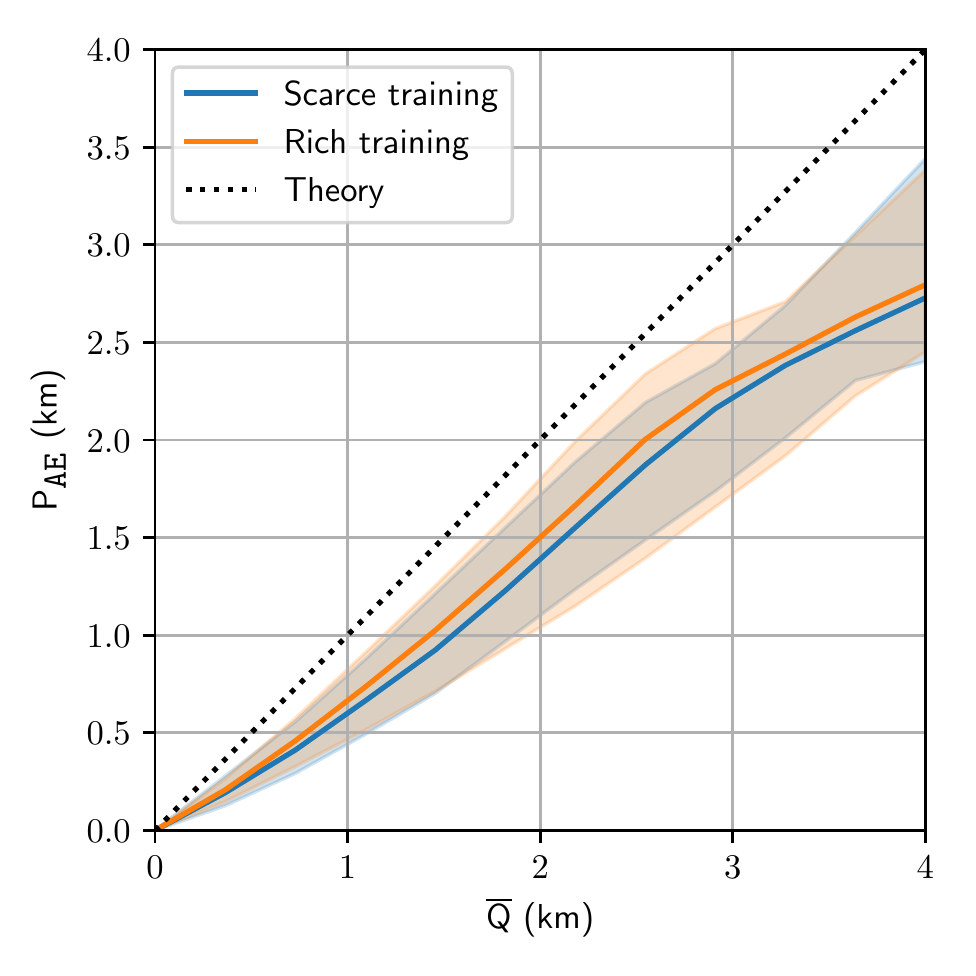}} 
  \hfil
  \subfloat[Brightkite, $\LHmk$.]{\includegraphics[width=0.25\linewidth]{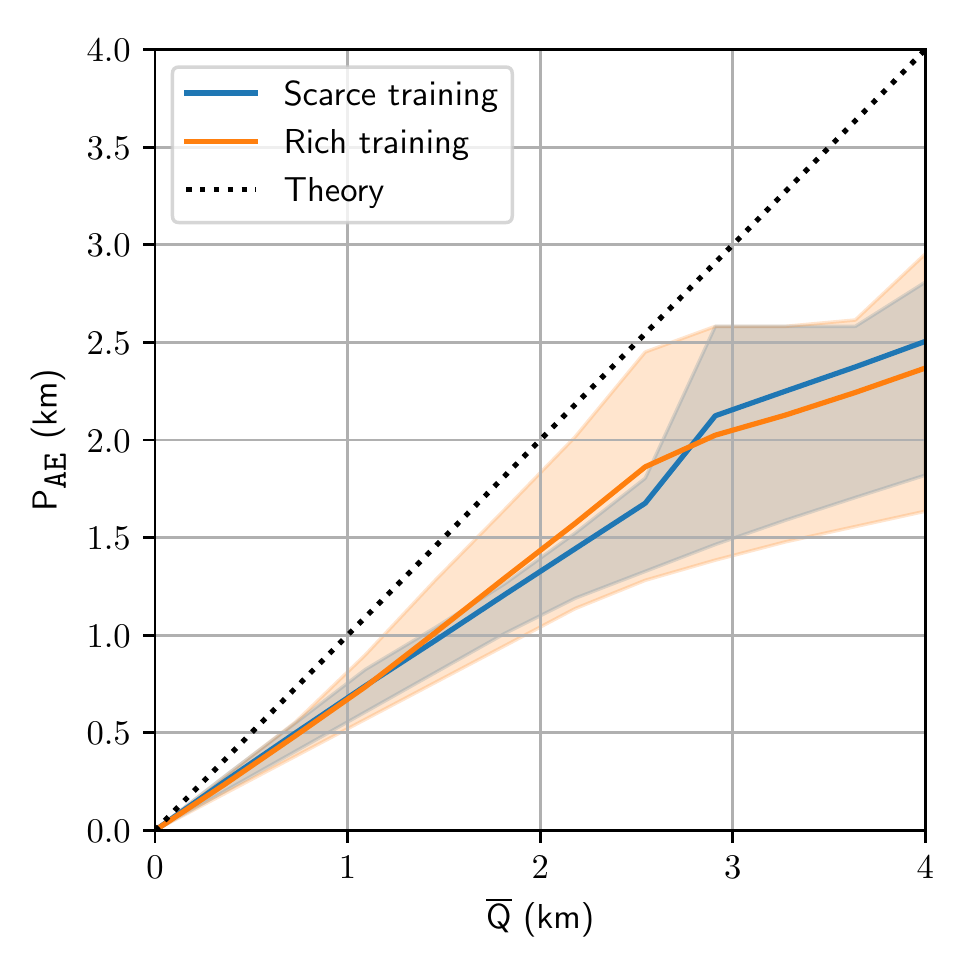}\label{fig:Eval1-Exp21-BKSP}}
  \hfil
  \subfloat[Gowalla, $\Expomk$.]{\includegraphics[width=0.25\linewidth]{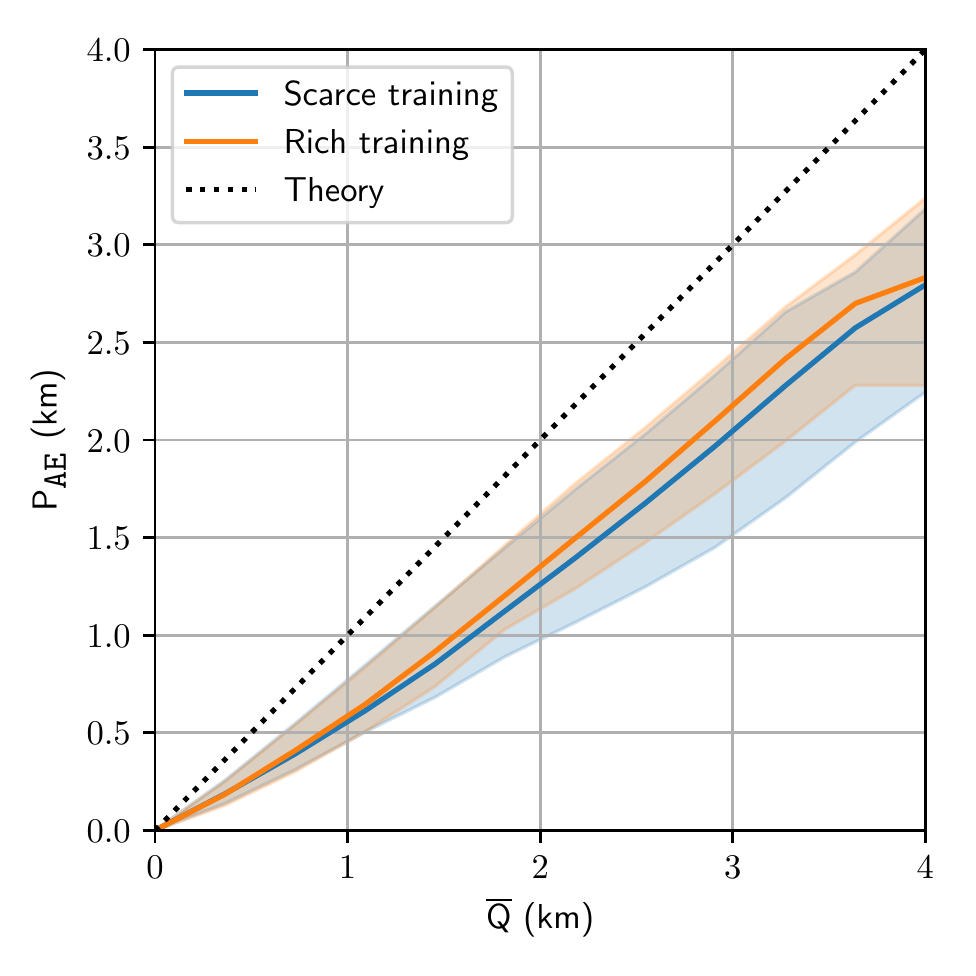}} 
  \hfil
  \subfloat[Gowalla, $\LHmk$.]{\includegraphics[width=0.25\linewidth]{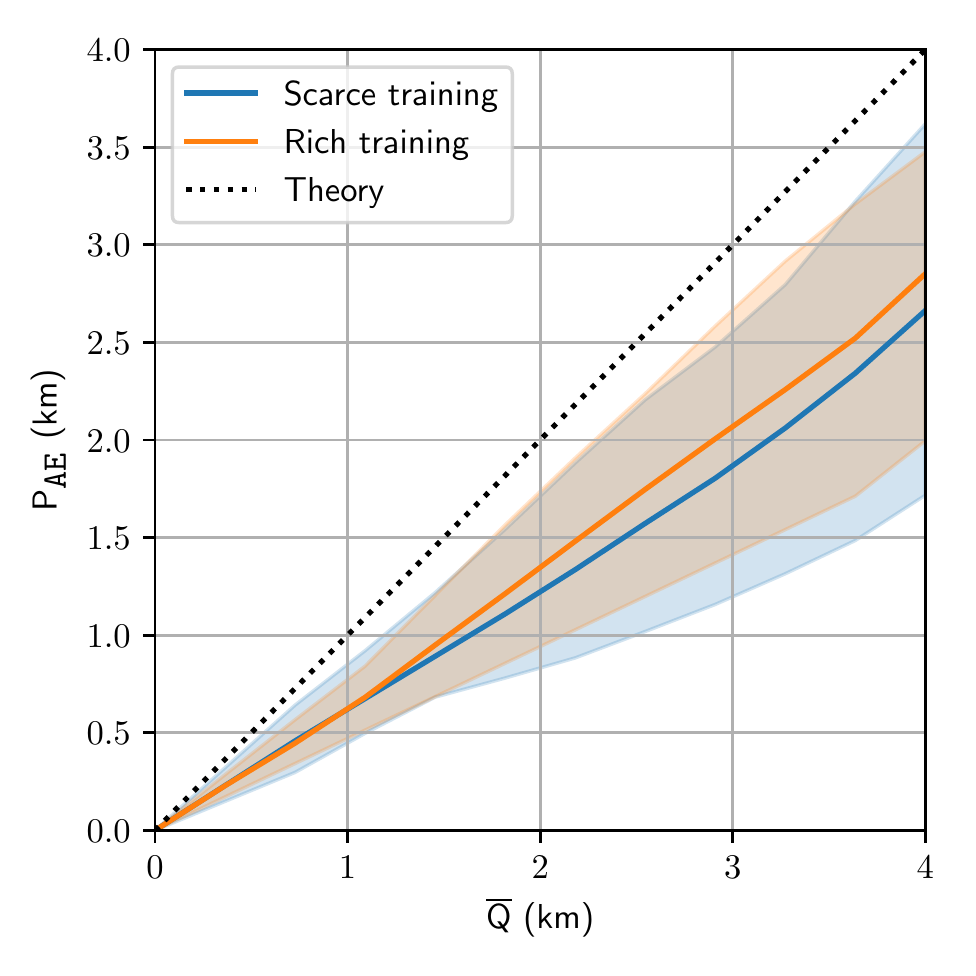}}
  \caption{Experiment MK: Performance of $\Expomk$ and $\LHmk$, against the optimal Markov attack in Brightkite and Gowalla datasets.}
  \label{fig:Eval1-Exp21}
\end{figure*}

\begin{figure} 
  \centering
  \subfloat[TaxiCab, $\Expomk$.]{\includegraphics[width=0.5\linewidth]{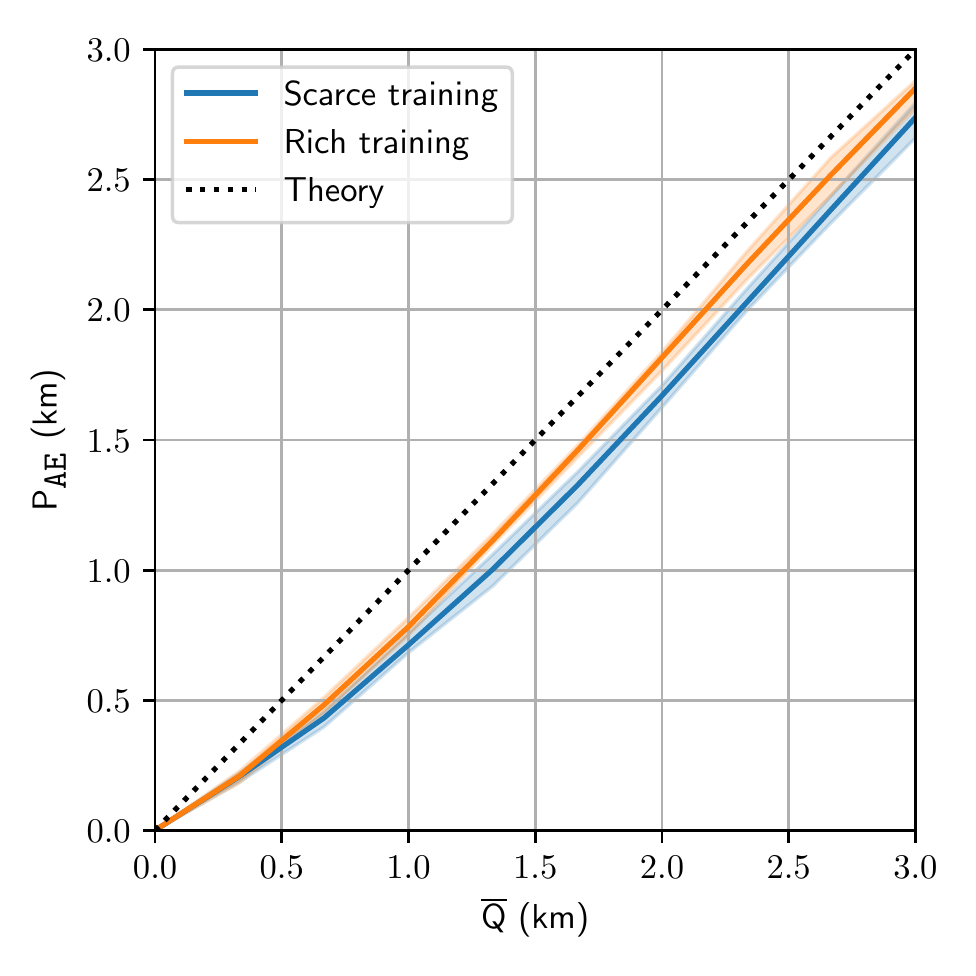}} 
  \hfil
  \subfloat[TaxiCab, $\LHmk$.]{\includegraphics[width=0.5\linewidth]{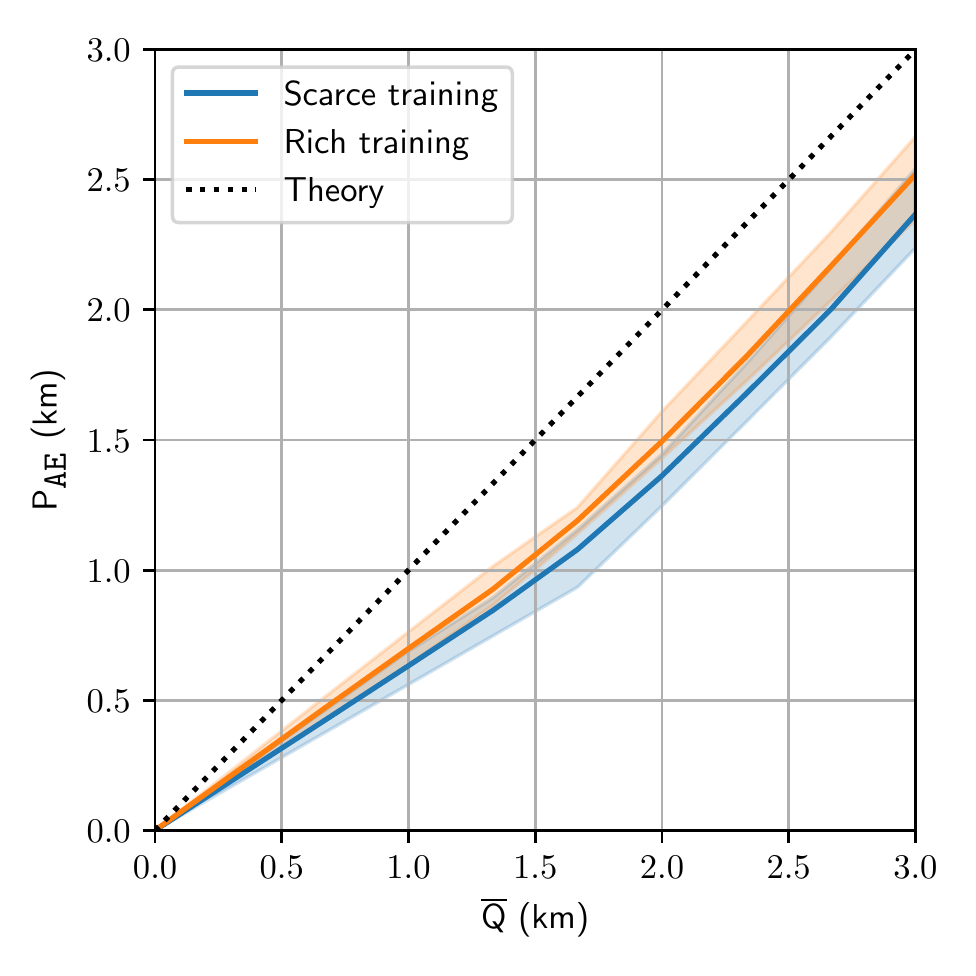}}
  \caption{Experiment MK: Performance of $\Expomk$ and $\LHmk$ against the optimal Markov attack in Taxicab dataset.}
  \label{fig:Eval1-Exp22}
\end{figure}

We evaluate $\LHmk$ and $\Expomk$ against the optimal Markov adversary. Figure~\ref{fig:Eval1-Exp21} shows the performance in Brightkite and Gowalla, and Fig.~\ref{fig:Eval1-Exp22} shows the performance in TaxiCab dataset. The results in Brightkite and Gowalla are very similar to the ones in the previous experiment, i.e., an optimal Markov LPPM that has been designed by hardwiring it on training data from other users provides significantly less privacy than expected in theory.

The results in TaxiCab dataset, however, are significantly better for the users. This is because, in this dataset, we have continuous location data (i.e., one location reported every 5 minutes). This means that two consecutive locations are highly correlated because of road restrictions (e.g., one-way roads, mandatory turns, etc.). Cabs follow very different paths each day, and thus it would seem that training their LPPMs with past data should not be significantly beneficial for them. However, the training data encodes these road restrictions. This is very important: the optimal Markov LPPMs are thus designed taking these constraints into account. Since the road restrictions are also part of the testing data, the optimal Markov LPPM is able to get close to optimal performance during evaluation. We also observe that training with seven days of data (rich training) is slightly better than training with a single day (scarce training). This slight improvement suggests that a single day of training already encodes most of the road restrictions. 
To validate this hypothesis, we also conducted experiments where we train each user's LPPM with past location traces of different users, and the results were similar (see Appendix~\ref{sec:appMK}). This confirms that taking road constraints into account is of paramount importance towards achieving high protection in continuous non-sporadic location privacy.

Finally, $\Expomk$ performs better than $\LHmk$. These results support the findings in~\cite{oya2017back}, where authors showed that exponential mechanisms perform significantly better than location hiding techniques when evaluated with metrics they have not been optimized for. In our case, we see that $\Expomk$ performs better than $\LHmk$ when evaluated in testing data it has not been optimized for.

\section{Blank-Slate Models}
\label{sec:blankslate}

We have seen that hardwiring the training data into the mobility models used for LPPM design can be detrimental to privacy. To alleviate this issue, we propose \emph{blank-slate} models for user mobility. These models treat their parameters ($\pi$ in the sporadic case; or $\pi_0$ and $M$ in the Markov case) as unknown variables that are never completely known to the user when designing her LPPM. These parameters can be initialized a-priori with training data, but do not remain fixed. Instead, the user updates them a-posteriori, as she acquires additional information from the observations (e.g., $\mathbf{x}$ and $\mathbf{z}$, from the testing set). Therefore, we can expect that LPPMs developed with blank-slate models will be desirable in situations where the training data does not adequately capture the user's mobility traits, either because it does not contain sufficient information or because it captures mobility patterns that are not characteristic of the user in question.

There are many ways in which a user can implement a blank-slate model. For example, a user can train a distribution on the hidden parameter (e.g., $p(\pi)$) based on training data, and then estimate this parameter a-posteriori using $\mathbf{x}$ and $\mathbf{z}$ (e.g., a maximum a-posteriori approach). In our case, we take a maximum likelihood approach that we explain below. We present a new family of LPPMs, the Profile Estimation-Based LPPMs (PEB-LPPMs), that we build by leveraging a \emph{sporadic blank-slate model} for user mobility. We do not tackle the problem of LPPM design under blank-slate Markov mobility models, but we show that our PEB-LPPM is also useful for users whose mobility model is Markovian.

\subsection{LPPM Design in the Sporadic Blank-Slate Model}
\label{sec:blankslate_design}

A sporadic blank-slate model is characterized by a mobility profile $\pi$ that is unknown to the user. In order to design an LPPM using this model, the user must first estimate this mobility profile. We propose to use a Maximum Likelihood Estimator (MLE) of the mobility profile before each query $r$, and then use this profile to build an optimal sporadic LPPM. We call the LPPM designed this way Profile Estimation Based (PEB)-LPPM.

More precisely, a PEB-LPPM is an output-based defense, i.e., characterized by $f(z^r|\mathbf{z}^{r-1},x^r)$, that is computed by following these steps:
\begin{enumerate}
 \item Compute an MLE of the mobility profile $\pi$ using $\mathbf{z}^{r-1}$. Let this estimate be $\hat{\pi}^r_{ML}$.
 \item Normalize the estimate $\hat{\pi}^r_{ML}$ to avoid variance issues for low $r$, producing $\hat{\pi}^r$.
 \item Compute the optimal LPPM in the sporadic mobility model using $\hat{\pi}^r$ and generate $z^r$ randomly using $x^r$ and this newly created LPPM.
\end{enumerate}
This whole process can be embedded into a function of the form $f(z^r|\mathbf{z}^{r-1},x^r)$ that defines the PEB-LPPM.

Note that, in the first step above, the user could have also used her past real locations $\mathbf{x}^{r-1}$ to compute the MLE estimation of her mobility profile (since she knows them). This, effectively, would result on a full-LPPM $f(z^r|\mathbf{z}^{r-1},\mathbf{x}^r)$. As we mentioned in Section~\ref{sec:hardwired_design}, assessing the privacy of these LPPMs against an optimal adversary is computationally intractable. Thus, to avoid gaps in our evaluation, we only use $\mathbf{z}^{r-1}$ to compute our MLE of the mobility profile.
We delve into the steps of the PEB-LPPM design below.

\subsection{Step 1: Mobility Profile Estimation.}
We derive the Maximum Likelihood Estimator (MLE) of the mobility profile given $\mathbf{z}^r$. We use $\pi_i\equiv p(x=x_i)$ to denote the probability mass function defined by $\pi$. Let $\mathcal{P}$ be the set of all the possible mobility profiles, i.e., $\mathcal{P}\doteq\{\pi| \sum_{i=1}^{|\mathcal{X}|} \pi_i = 1,\quad \pi_i\geq 0\}$. The MLE of the mobility profile is the estimation $\hat{\pi}^r_{ML}$ that maximizes $p(\mathbf{z}^r|\pi)$, i.e.,
\begin{equation} \label{eq:MLE}
\hat{\pi}^r_{ML}= \underset{\pi\in\mathcal{P}}{\text{ argmax }} p(\mathbf{z}^r|\pi)\,. \\
\end{equation}

Solving this problem directly is complicated, since $\mathbf{z}^r$ is not generated directly from $\pi$, but from $\mathbf{x}^r$. We use the Expectation-Maximization (EM) method~\cite{agrawal2001design} to find an efficient way of computing this estimator. We rely on $\mathbf{x}^r$ as auxiliary data, and define the auxiliary function $Q$ as
\begin{equation}
 Q(\pi,\pi^t)\doteq\Exp{\log p(\mathbf{x}^r|\pi)|\mathbf{Z}=\mathbf{z}^r,\Pi=\pi^t}\,.
\end{equation}
The EM method iterates over two steps: first, compute $Q(\pi,\pi^t)$ (E-step), and then find $\pi^{t+1}$ as the profile $\pi$ that maximizes $Q(\pi,\pi^t)$ (M-step). We expand $Q$ as
\begin{equation}
\begin{split}
 Q(\pi,\pi^t)&\doteq\Exp{\log p(\mathbf{x}^r|\pi)|\mathbf{Z}^r=\mathbf{z}^r,\Pi=\pi^t}\\
  &=\sum_{s=1}^r \Exp{\log p(x^s|\pi)|\mathbf{Z}^r=\mathbf{z}^r,\Pi=\pi^t}\\
  &=\sum_{s=1}^r \sum_{i=1}^{|\mathcal{X}|} \log \pi_i \cdot p(x^r_i|\mathbf{z}^r,\pi^t)\\
  &=\sum_{i=1}^{|\mathcal{X}|} \log \pi_i \cdot \left[\sum_{s=1}^r  p(x^s_i|\mathbf{z}^r,\pi^t)\right]\,.
\end{split}
\end{equation}

In order to find the $\pi\in\mathcal{P}$ that maximizes $Q(\pi,\pi^t)$, we build the Lagrange multipliers function
\begin{equation}
 L(\pi,\lambda, \boldsymbol{\mu})=Q(\pi,\pi^t)+\lambda\left(\sum_{i=1}^{|\mathcal{X}|}\pi_i-1\right)+\sum_{i=1}^{|\mathcal{X}|} \mu_i \pi_i\,,
\end{equation}
where the term with $\lambda$ corresponds to the constraint $\sum_{i=1}^{|\mathcal{X}|}\pi_i=1$ and the terms with $\mu_i$ correspond to $\pi_i\geq 0$. We take $\mu_i=0$ for the non-negativity constraints, and by solving $\partial L/\partial \pi_i=0$ and $\partial L/\partial \lambda=0$ we obtain the maximum, which gives us the update rule $\pi_i^{t+1}=\frac{1}{r} \sum_{s=1}^r p(x^s_i|\mathbf{z}^s,\pi^t)$. We use Bayes' Rule to expand $p(x_i^s|\mathbf{z}^s,\pi^t)$ and we finally obtain
\begin{equation} \label{eq:updatepi}
 \pi_i^{t+1}=\frac{1}{r} \sum_{s=1}^r \frac{\pi^t_i \cdot f(z^s|\mathbf{z}^{s-1},x^s_i)}{\sum_{k=1}^{|\mathcal{X}|} \pi^t_k \cdot f(z^s|\mathbf{z}^{s-1},x^s_k)}\,.
\end{equation}
Following~\cite{boyd2004convex}, we can see that this solution is the global maximum of $Q(\pi,\pi^t)$, since it meets the KKT conditions, $Q(\pi,\pi^t)$ is strictly concave on $\pi$ (it is a weighted sum of logarithms) and $\mathcal{P}$ is a convex set.

Summarizing, in order to compute the MLE of the mobility profile, one proceeds as follows. First, define an initial profile $\pi^0$. Then, follow the update rule given by \eqref{eq:updatepi} until convergence (i.e., until the change from $\pi^t$ to $\pi^{t+1}$ is small enough). This algorithm is ensured to converge to the MLE for memoryless and output-based LPPMs, as we prove in Appendix~\ref{sec:proofEM}. 

\subsection{Step 2: MLE Normalization}

The accuracy of the MLE estimator above depends on the number of queries done previously. For example, we can expect to have a worse estimation of $\pi$ if we compute it at time $r=2$ using only $z^1$, compared to computing it at time $r=100$ with $\mathbf{z}^{99}$. To alleviate this issue, we perform a normalization step. Let $\pi_{ini}$ be an \emph{initial} mobility profile (e.g., a uniform profile, a profile computed from auxiliary data, or a profile computed from the training data as in hardwired models), and let $\gamma>0$ be a constant. The final mobility profile after the normalization step is
\begin{equation} \label{eq:pinorm}
 \hat{\pi}^r=\frac{1}{r^{\gamma}}\cdot \pi_{ini} + \left(1-\frac{1}{r^{\gamma}}\right)\cdot \hat{\pi}_{ML}^r\,.
\end{equation}
The coefficient $\gamma$ tunes how fast the effect of $\pi_{ini}$ in $\hat{\pi}^r$ fades with $r$. For example, if the user does not have enough data to compute a reliable initial profile $\pi_{ini}$, she can simply set $\gamma=0.5$ so that $\hat{\pi}^r$ converges fast to the ML estimate $\hat{\pi}_{ML}^r$. If the user believes that $\pi_{ini}$ is representative of her current mobility behavior, a slower rate $\gamma=0.1$ is more appropriate. 

\subsection{Step 3: Final LPPM Computation}
\label{sec:blankslate_design_3}
  
Once the user has computed her estimation of the mobility profile $\hat{\pi}^r$ she builds an optimal memoryless LPPM for the sporadic location privacy case (e.g., using the linear programming or the optimal remapping approach we explained in Section~\ref{sec:hardwired_design_sporadic}). Using this LPPM, she samples the obfuscated location $z^r$ given her real location $x^r$.

\section{Evaluation of Profile Estimation-Based LPPMs}
\label{sec:eval2}

We evaluate the performance of the PEB-LPPMs that we developed with the blank-slate sporadic mobility model, and compare them with the optimal LPPMs that we evaluated earlier. We use the notation $\LHbs$ and $\Expobs$ to denote the location hiding and exponential LPPMs computed following the PEB-LPPM strategy in Sect.~\ref{sec:blankslate}. We compute $\pi_{ini}$ from the training set and chose $\gamma=0.5$ so that the PEB-LPPMs adapt quickly to the MLE of the mobility profile. This means that, after $r=100$ queries, the mobility profile that is used for design $\hat{\pi}^r$ in \eqref{eq:pinorm} is $\hat{\pi}^r=0.1\cdot \pi_{ini} + 0.9\cdot \hat{\pi}_{ML}^r$. 

We split the evaluation into two parts, using the same settings of Section~\ref{sec:eval1} (see Table~\ref{tab:experiments1}). Since PEB-LPPMs learn the user behavior as she queries the LBS, we can expect that their performance will improve over time. Therefore, instead of averaging $\Qavg(f,r)$ and $\PAE(f,h,r,r)$ over all values of $r$, we perform the average over the first and last halves separately (e.g., $\sum_{r=1}^{150}\dots$ and $\sum_{r=151}^{300}\dots$ in Brightkite/Gowalla). The performance of PEB-LPPMs averaged over the last half of the location releases is practically independent of the initial profile $\pi_{ini}$ since, for $r>150$, \eqref{eq:pinorm} becomes $\hat{\pi}^r\approx \hat{\pi}_{ML}^r$.

\subsection{Experiment SP with PEB-LPPMs}

\begin{figure*} 
  \centering
  \subfloat[Brightkite, $\Expo$.]{\includegraphics[width=0.25\linewidth]{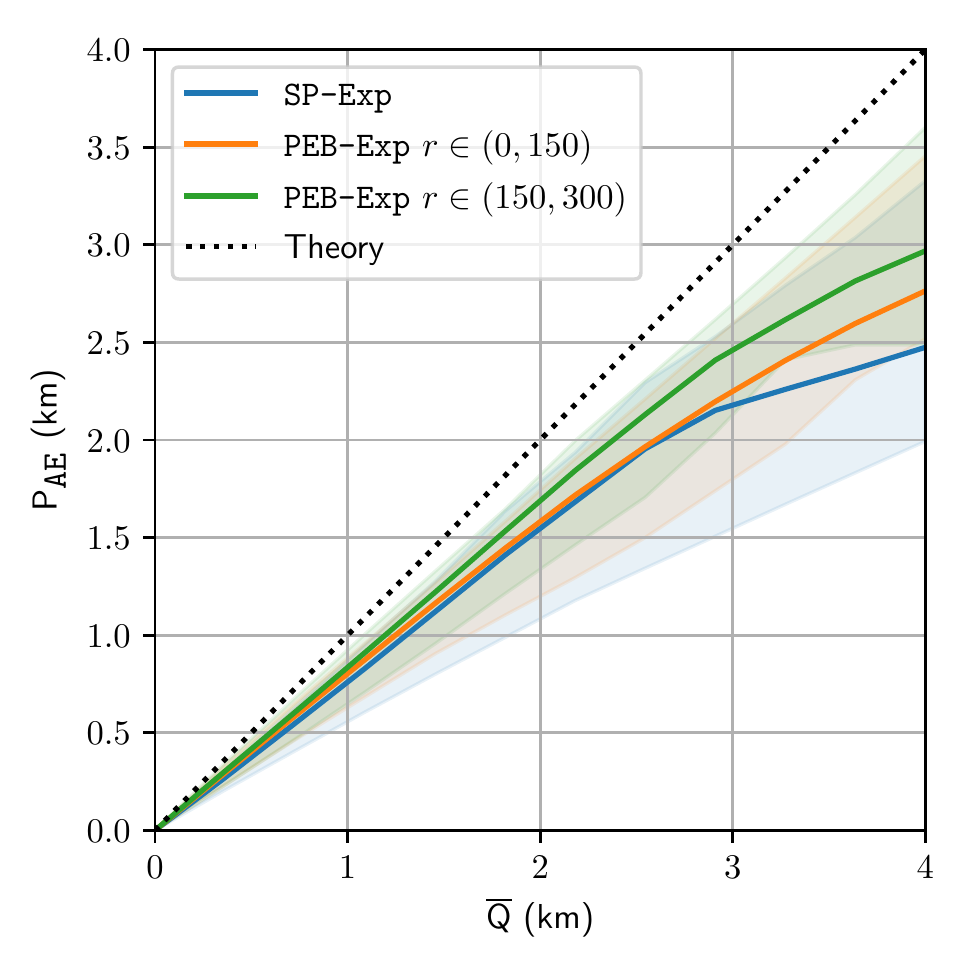}} 
  \hfil
  \subfloat[Brightkite, $\LH$.]{\includegraphics[width=0.25\linewidth]{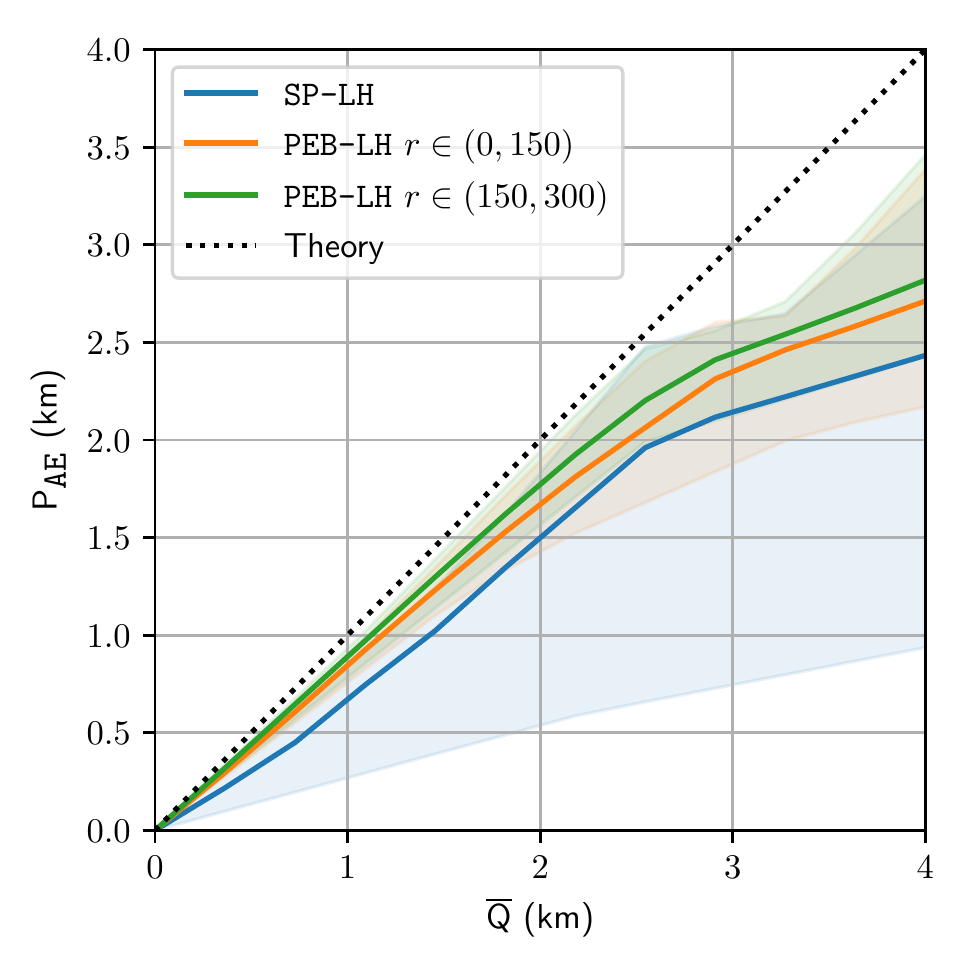}}
  \hfil
  \subfloat[Gowalla, $\Expo$.]{\includegraphics[width=0.25\linewidth]{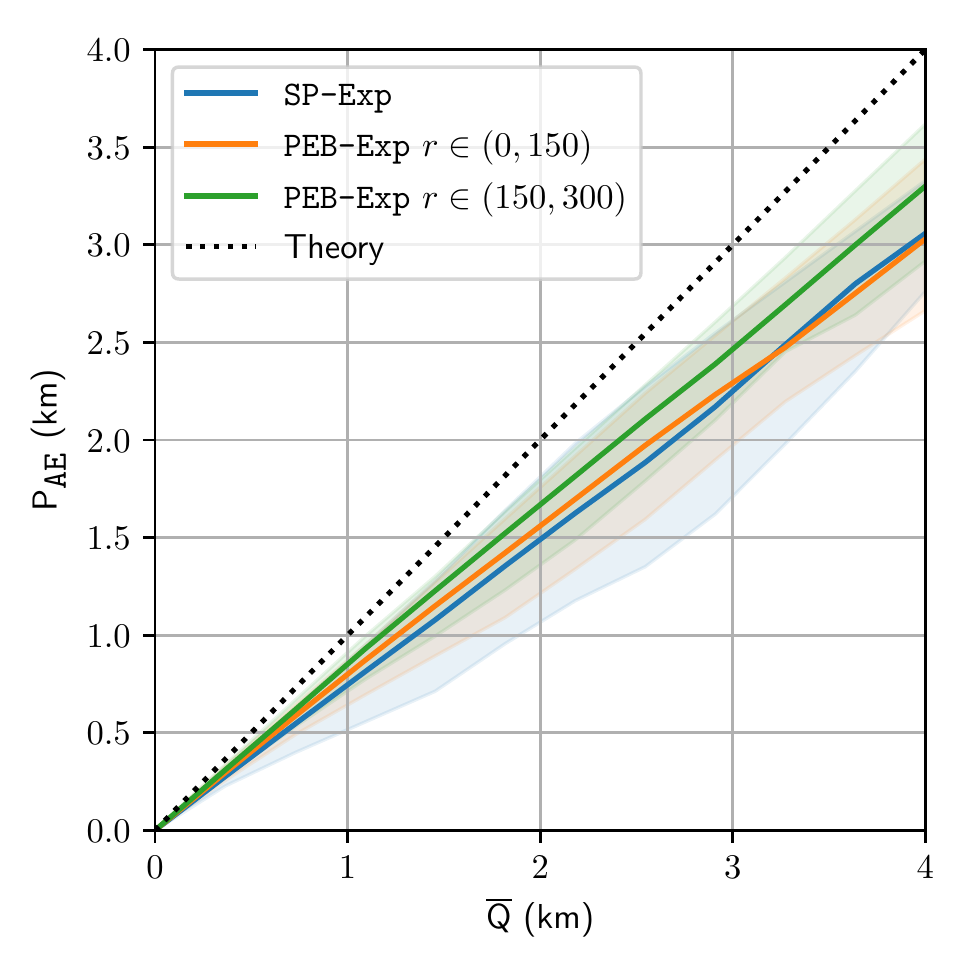}} 
  \hfil
  \subfloat[Gowalla, $\LH$.]{\includegraphics[width=0.25\linewidth]{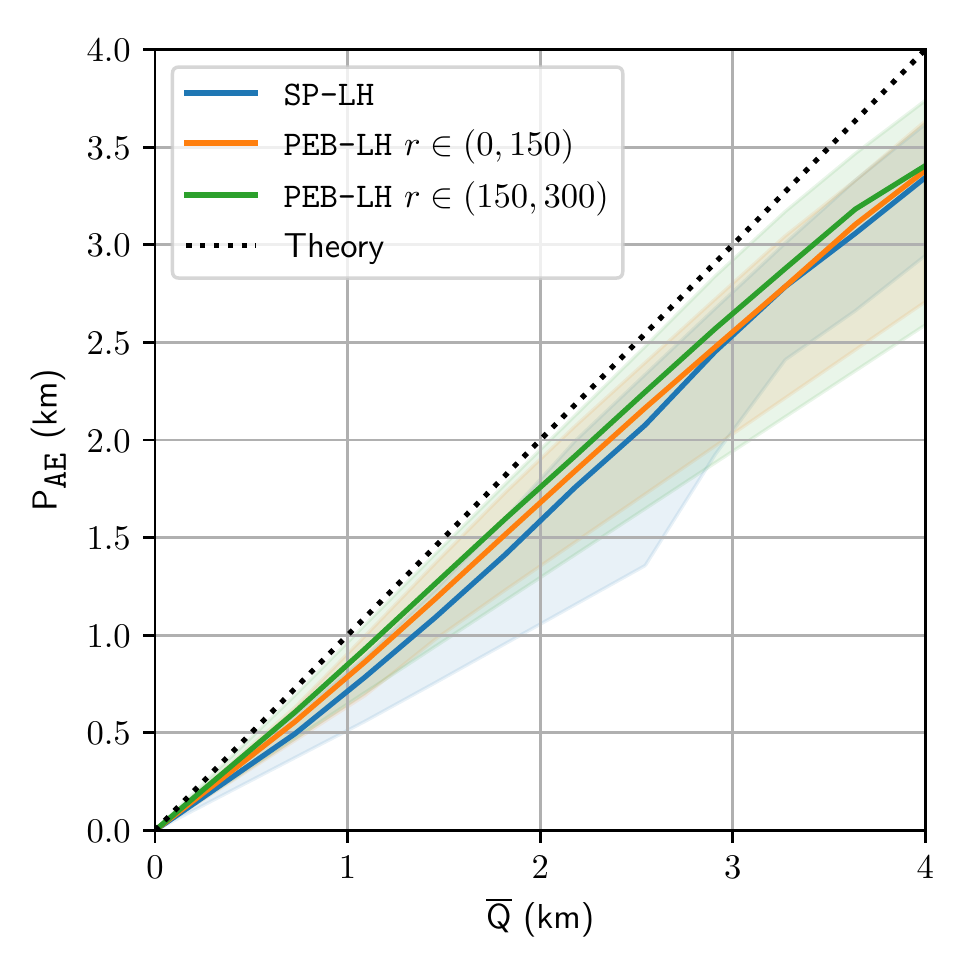}}
  \caption{Experiment SP: Performance of PEB-LPPMs versus $\HWsporadic$, using Brightkite and Gowalla datasets (shuffled). $\HWsporadic$ have been trained with the rich data.}
  \label{fig:Eval2-Exp1}
\end{figure*}

First, we evaluate PEB-LPPMs in the sporadic scenario. We compare the performance of $\LHbs$ and $\Expobs$ with $\LHsp$ and $\Exposp$, against the optimal sporadic adversary. We use only the rich data to train $\LHsp$ and $\Exposp$ and to initialize $\LHbs$ and $\Expobs$ (for simplicity). Figure~\ref{fig:Eval2-Exp1} shows the results. The blue line corresponds to the orange line in Fig.~\ref{fig:Eval1-Exp1} ($\HWsporadic$ trained with the rich data). The orange line is the average performance of PEB-LPPMs in the first 150 samples, and the green line is the average performance in last 150 samples. We can see that PEB-LPPMs always outperform hardwired ones ($\HWsporadic$) in the sporadic scenario, and that the performance of PEB-LPPMs improves with $r$. This is reasonable, as these mechanisms estimate the real user behavior adaptively during the evaluation, and with higher $r$ values this estimation is more accurate. These results show that disregarding the training data and relying solely on the MLE of the mobility profile ($\LHbs$ and $\Expobs$ with $r>150$) can yield LPPMs that offer better protection than those hardwired on the training data ($\LHsp$ and $\Exposp$).

\subsection{Experiment MK with PEB-LPPMs}

\begin{figure*} 
  \centering
  \subfloat[Brightkite, $\Expo$.]{\includegraphics[width=0.25\linewidth]{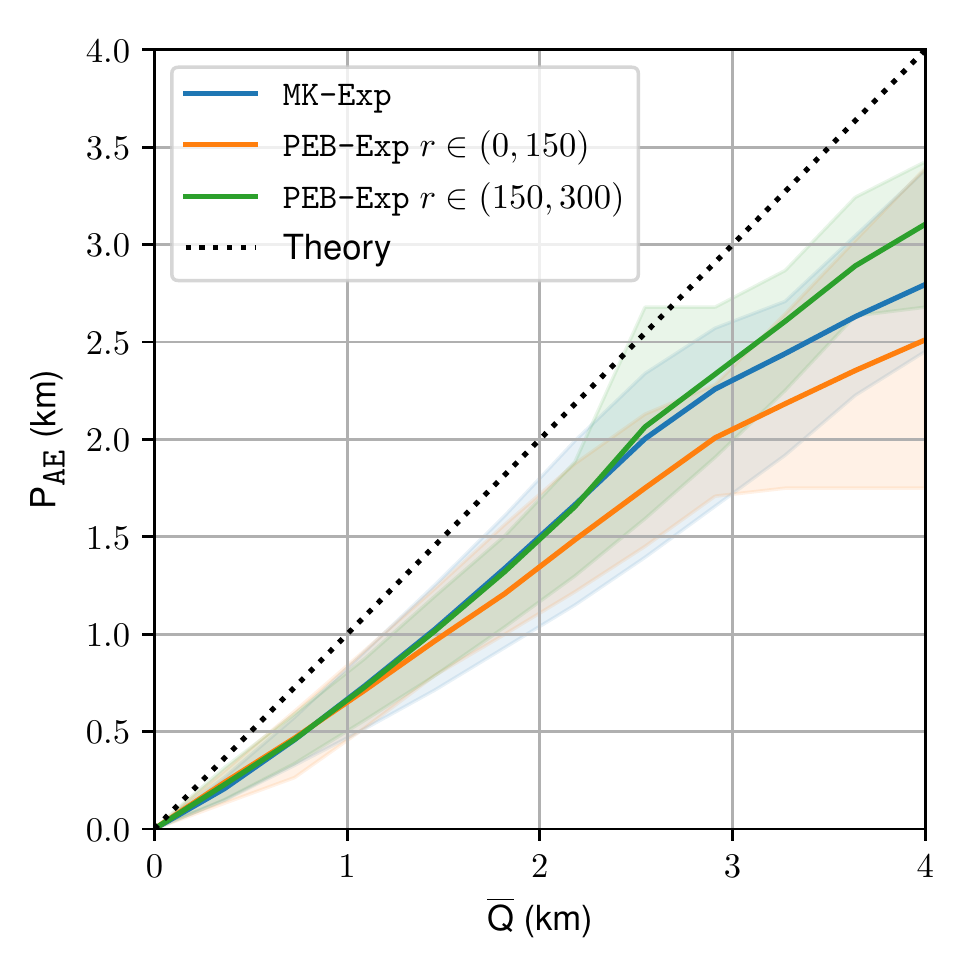}} 
  \hfil
  \subfloat[Brightkite, $\LH$.]{\includegraphics[width=0.25\linewidth]{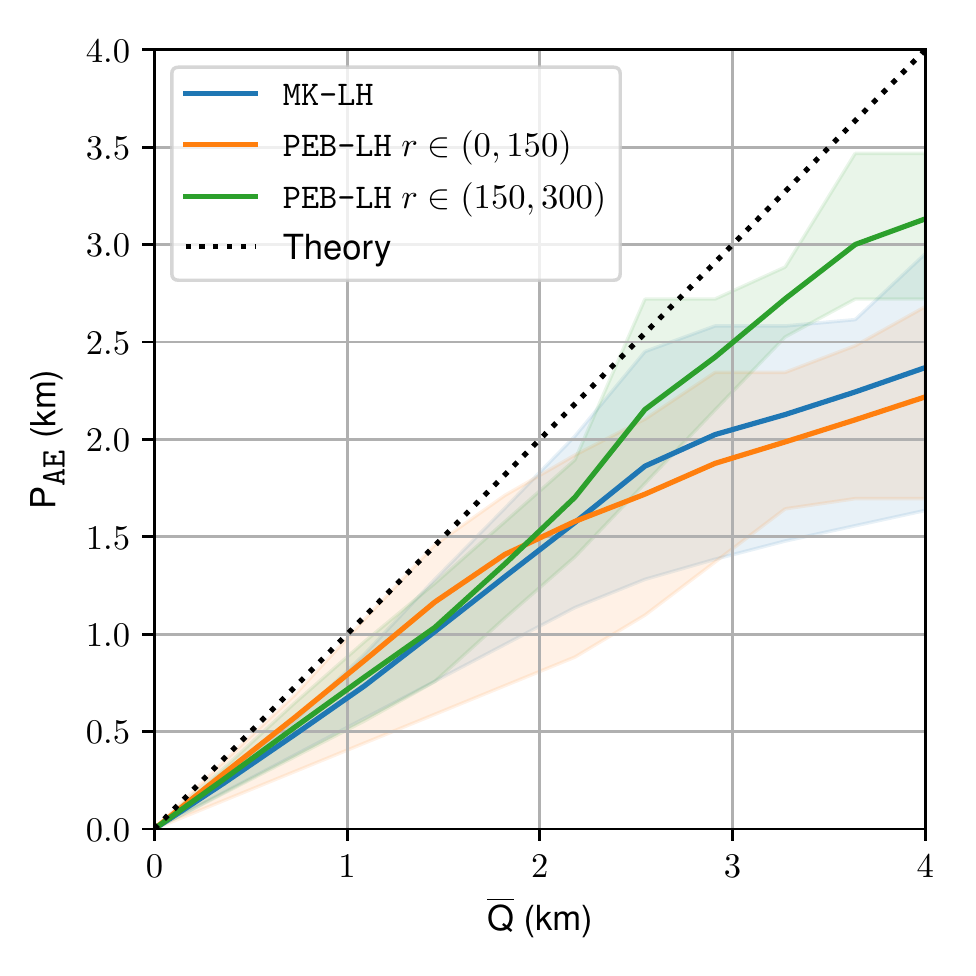}}
  \hfil
  \subfloat[Gowalla, $\Expo$.]{\includegraphics[width=0.25\linewidth]{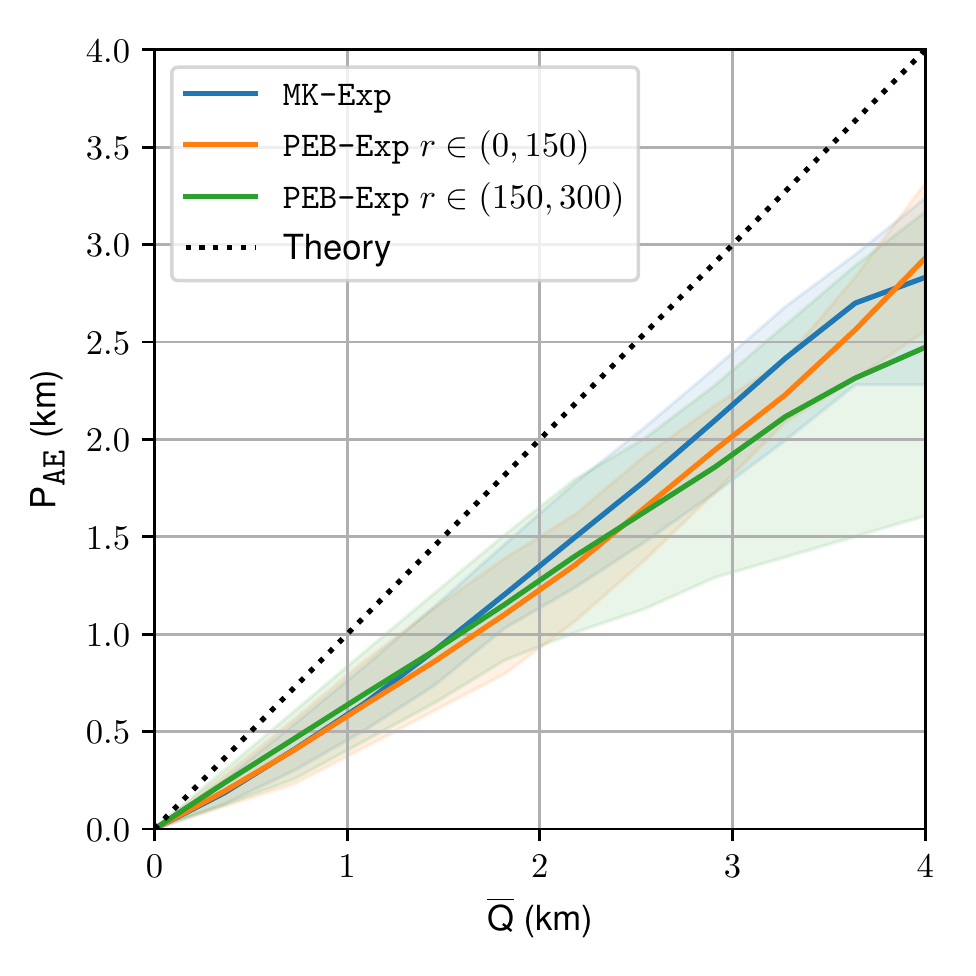}} 
  \hfil
  \subfloat[Gowalla, $\LH$.]{\includegraphics[width=0.25\linewidth]{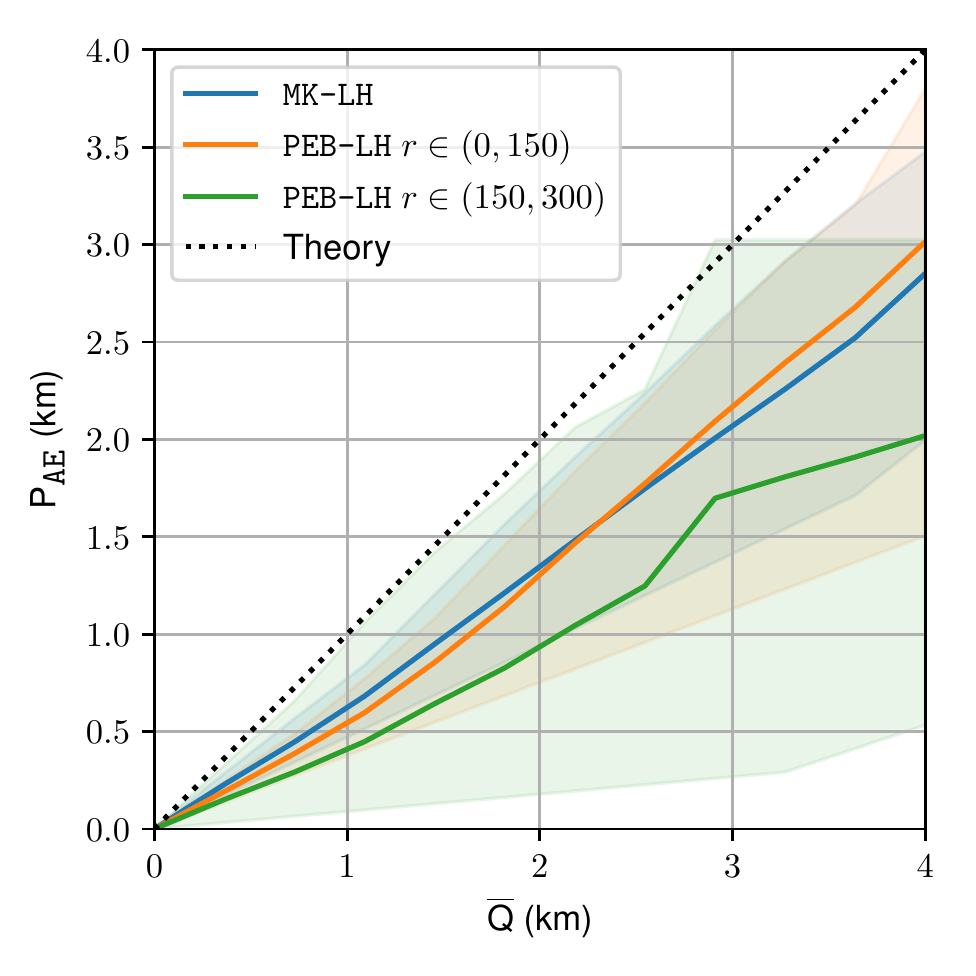}}
  \caption{Experiment MK: Performance of PEB-LPPMs versus $\HWmarkov$ in Brightkite and Gowalla datasets. $\HWmarkov$ have been trained with the rich data.}
  \label{fig:Eval2-Exp21}
\end{figure*}

\begin{figure} 
  \centering
  \subfloat[TaxiCab, $\Expo$.]{\includegraphics[width=0.50\linewidth]{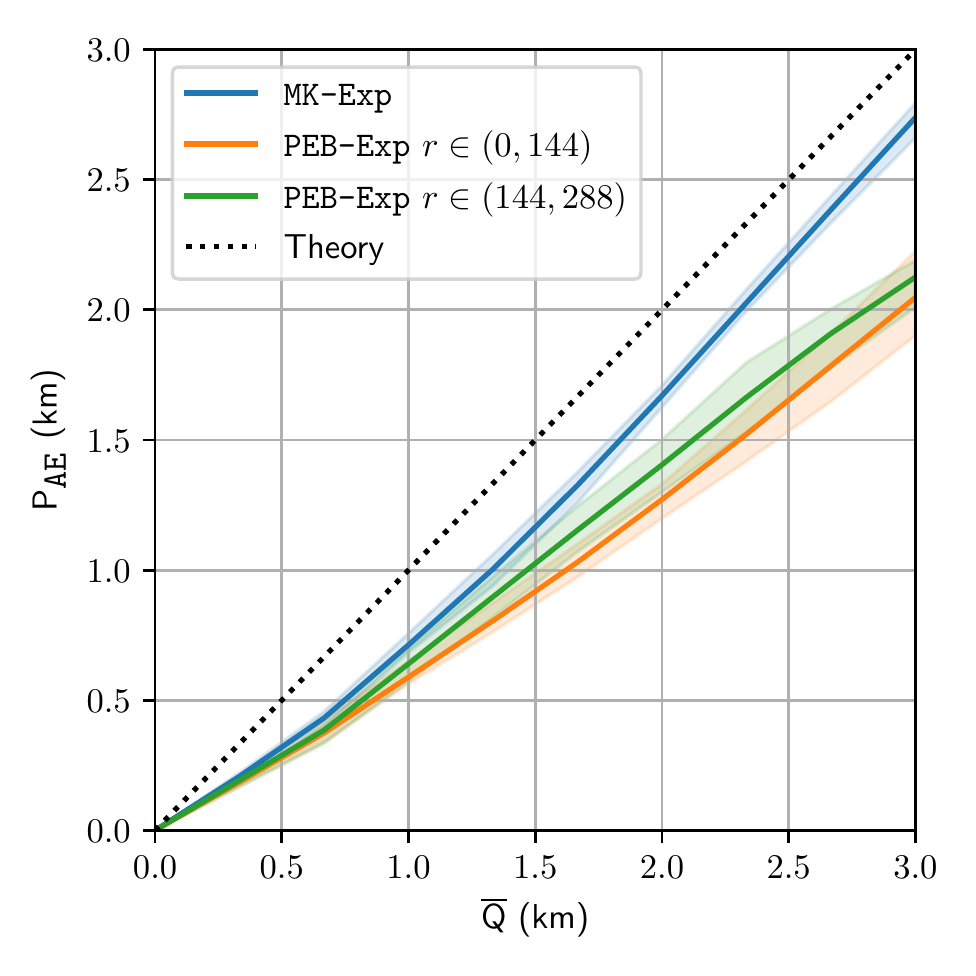}} 
  \hfil
  \subfloat[TaxiCab, $\LH$.]{\includegraphics[width=0.50\linewidth]{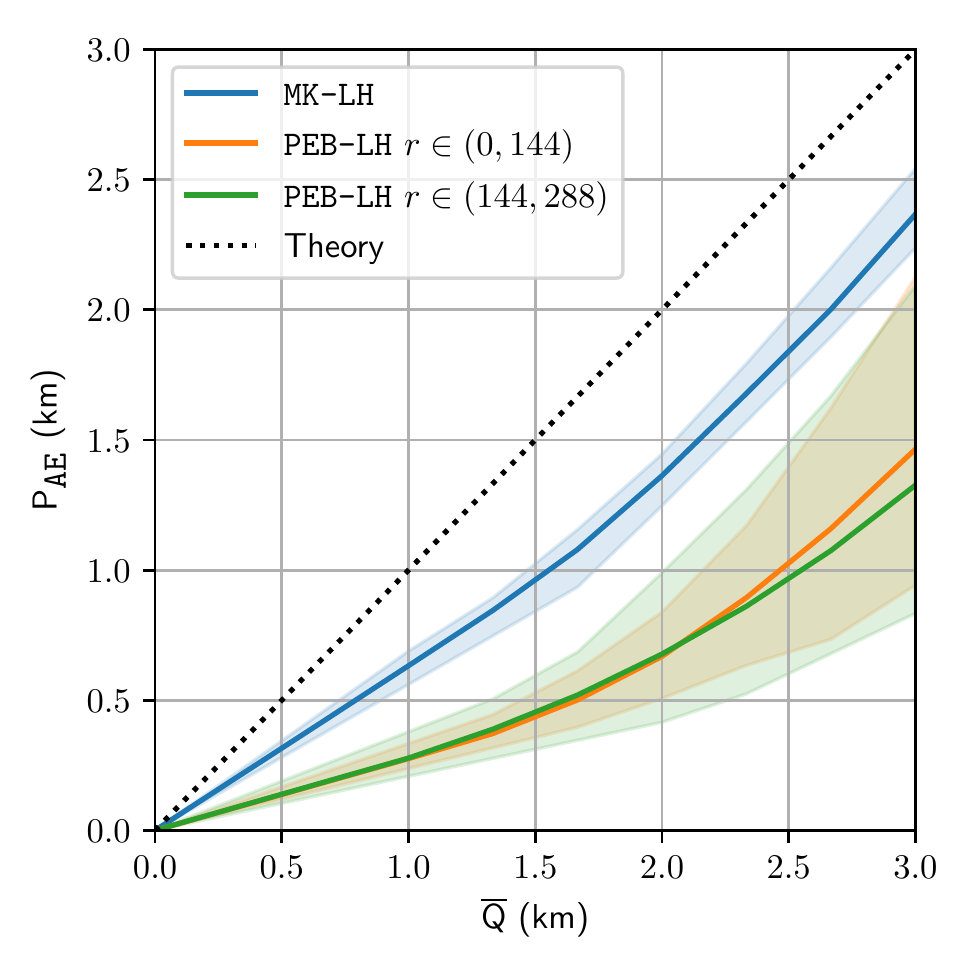}} 
  \caption{Experiment MK: Performance of PEB-LPPMs versus $\HWmarkov$ in TaxiCab dataset, with one day of training. $\HWmarkov$ have been trained with the scarce data (a single-day trace).}
  \label{fig:Eval2-Exp22}
\end{figure}

Now, we compare $\LHbs$ and $\Expobs$ with $\LHmk$ and $\Expomk$ against the optimal Markov adversary, in the settings of Experiment MK. Figure~\ref{fig:Eval2-Exp21} shows the results for Brightkite and Gowalla, and Fig.~\ref{fig:Eval2-Exp22} for TaxiCab. In Brightkite and Gowalla, we use only the rich training data to build $\HWmarkov$. Here, even though PEB-LPPMs are built upon the sporadic blank-slate mobility model, they are on-par with optimal Markov designs in non-sporadic location privacy settings, and in many cases outperform them. This is because Brightkite and Gowalla are datasets where user check-ins are not strongly correlated. This means that capturing the road restrictions is not decisive towards achieving a good privacy performance, and therefore PEB-LPPMs can compete with $\HWmarkov$. 
Note that the performance of PEB-LPPMs decreases with $r$ in Gowalla. This is due to the fact that the MLE of the mobility profile uses all the past observations and does not give more weight to recent ones. If the user behavior changes drastically at some point in the trace (e.g., around $r=150$), the MLE will not be able to follow this change, and thus the performance of the PEB-LPPM will decrease. It would be interesting to study how to adjust PEB-LPPMs so that they give more \emph{weight} to recent location releases when estimating the mobility profile of the user.

The situation changes in TaxiCab dataset (Fig.~\ref{fig:Eval2-Exp22}). In this case, even though we have decided to train $\HWmarkov$ using the scarce training set (one day of data for each user), this is enough for $\HWmarkov$ to achieve an outstanding performance (as we saw in Fig.~\ref{fig:Eval1-Exp22}). This is because, in TaxiCab dataset,  the locations are tightly correlated due to road restrictions. PEB-LPPMs are built leveraging a sporadic blank-slate model, so they cannot capture these restrictions, and thus perform poorly in this dataset. Note that increasing $r$ does not have a significant effect in the performance, since it does not matter how accurately the profile estimation of PEB-LPPMs is: a (sporadic) mobility profile cannot capture the correlations of non-sporadic models.

\subsection{Summary of Results and Other Privacy Metrics}
\label{sec:discussion}

PEB-LPPMs outperform optimal hardwired LPPMs in all of our \emph{sporadic} location privacy experiments. This is reasonable, as in these experiments the training data cannot closely characterize the behavior of the testing set users. This does not mean that PEB-LPPMs \emph{always} outperform hardwired LPPMs in sporadic location release scenarios: if user behavior can be accurately modeled by the training data, the performance of hardwired LPPMs would be close to optimal. However, we can confirm that PEB-LPPMs are a powerful tool to protect users whose mobility behavior cannot be predicted from the training data.


In \emph{non-sporadic} location privacy, our experiments show that PEB-LPPMs can outperform optimal Markov LPPMs when the user's real locations are not highly correlated (i.e., Brightkite and Gowalla datasets). When there are high dependencies between the real locations (i.e., TaxiCab data with location reports every 5 minutes), PEB-LPPMs perform worse than optimal Markov designs because they cannot capture these correlations. This could be addressed in future work by developing PEB-LPPMs based on \emph{blank-slate Markov} models. These PEB-LPPMs would re-estimate the Markov transition matrix on-the-fly using released locations and taking road restrictions into account.


On another note, in this work we use the average adversary error to measure privacy, since it is probably the most used metric in related works~\cite{Quant2011SP,Opt2012CCS,GeoInd2013CCS,OptGeoInd2014CCS,Elastic2015PETS,oya2017back,Semantics2016PETS,theodorakopoulos2014prolonging,shokri2017privacy}. However, there are other alternatives, such as geo-indistinguishability~\cite{GeoInd2013CCS,chatzikokolakis2014predictive,OptGeoInd2014CCS,OyaTP17,chatzikokolakis2017efficient} or the conditional entropy~\cite{Quant2011SP,oya2017back}, that quantify different notions of privacy. We now discuss why PEB-LPPMs also improve in terms of these metrics.

Throughout our evaluation (Figs.~\ref{fig:Eval2-Exp1}-\ref{fig:Eval2-Exp22}) we have seen that, in many scenarios, PEB-LPPMs outperform hardwired-based LPPMs. The underlying reason of this improvement is that the mobility profile that PEB-LPPMs estimate a-posteriori characterizes the actual user mobility better than the hardwired models. Thus, we can expect that PEB-LPPMs will also outperform hardwired LPPMs in these scenarios in terms of other privacy metrics, since they are more tailored to the actual user behavior in the testing data (e.g., $\Expobs$ will provide more geo-indistinguishability or conditional entropy than $\Exposp$ for the same quality loss, in a sporadic location release scenario). 

\section{Related Work}
\label{sec:relwork}

Early surveys of location privacy attacks, defenses and privacy metrics, by Decker~\cite{decker2008location} and Krumm~\cite{krumm2009survey}, do not include any discussion about modeling user mobility.

A first explicit modeling appears in~\cite{Quant2011SP}, where Shokri et al. propose a framework to evaluate location privacy mechanisms. In their framework instantiation, they consider a Markov hardwired model for user mobility, and in their evaluation they effectively merge training and testing sets.
A number of follow-ups also hardwire the mobility model using the evaluation data itself. In sporadic location privacy, this methodology was used to design and evaluate LPPMs according to different privacy notions. First, it was used to find optimal LPPMs in terms of the average adversary error, either by reporting individual locations~\cite{Opt2012CCS}, using dummy check-ins~\cite{herrmann2013optimal}, or in combination with geo-ind guarantees~\cite{Shokri15}. Second, hardwired user mobility models are used to obtain utility improvements and derive optimal geo-indistinguishability LPPMs~\cite{OptGeoInd2014CCS}, or to evaluate a semantic variation of this notion~\cite{Elastic2015PETS}.
Non-sporadic location privacy works also hardwired their mobility models on the evaluation data, and typically adopt a Markov model for user mobility to account for temporal correlations~\cite{Semantics2016PETS,shokri2017privacy}.

Chatzikokolakis et al. are the first to explicitly separate data used to design LPPMs and to evaluate them~\cite{chatzikokolakis2017efficient}, in the context of geo-indistinguishability. However, they do not quantify the privacy gap between theoretical design and empirical evaluation in a testing set. 

To the best of our knowledge, our work is the first to evaluate previous optimal LPPMs by considering a separation between training and testing data. We find out that LPPMs perform worse than previously reported results~\cite{Opt2012CCS,shokri2017privacy,oya2017back,theodorakopoulos2014prolonging} when empirically evaluated in a testing set. We also propose a blank-slate model for user mobility, which allows us to design LPPMs that \emph{learn} the model parameters during the evaluation. We are not aware of other blank-slate models in the literature, although the mobility profile estimation carried by PEB-LPPMs is similar to the problem of estimating a distribution from noisy data in privacy-preserving data mining~\cite{agrawal2001design}. 

\section{Conclusions}
\label{sec:conclusions}

Previous strategies to design Location Privacy-Preserving Mechanisms (LPPMs) assume that training data can completely characterize user mobility behavior, and hardwire this information in the mechanism itself. We demonstrate how this design decision overestimates the privacy offered by these designs when the users' mobility profile deviates from the training set characteristics.

We propose to use blank-slate models for user mobility that treat the mobility profile as an unknown variable that has to be \emph{learned}. We leverage a sporadic blank-slate model to propose a new family of defense techniques, PEB-LPPMs, that adapt to the user behavior using past obfuscated queries. We compare our proposal to hardwired LPPMs, and show that PEB-LPPMs improve the privacy except in continuous location release scenarios where user locations are highly correlated.

The problem identified in this paper is not unique to the location privacy domain. More generally, to build privacy enhancing technologies that provide strong privacy guarantees in real cases, we have to embrace that training information cannot always fully capture real user behavior. We believe that blank-slate models, that incorporate the uncertainty about real user behavior, are a promising approach to improve the protection provided by privacy mechanisms not only in location privacy but in a broader type of privacy problems.

\appendix

\subsection{Privacy performance of the memoryless LPPM in the hardwired model of user mobility.}
\label{sec:proofMemoryless}

Consider the full-type LPPM $f(z^r|\mathbf{z}^{r-1},\mathbf{x}^r)$, and a memoryless-type LPPM that we denote by $f^*$, defined as
\begin{equation}
 f^*(z^r|x^r)\doteq \sum_{\substack{\mathbf{x}^{r-1}\\ \in\mathcal{X}^{r-1}}}\sum_{\substack{\mathbf{z}^{r-1}\\ \in\mathcal{Z}^{r-1}}} p(\mathbf{x}^{r-1},\mathbf{z}^{r-1}|x^r) \cdot f(z^r|\mathbf{z}^{r-1},\mathbf{x}^r)\,.
\end{equation}

The average loss of $f$ and $f^*$ is the same, i.e., $\Qavg(f,r)=\Qavg(f^*,r)$ due to the linearity of this metric. Then, by proving that $f^*$ does not achieve less privacy than $f$, we prove that the privacy and quality loss trade-off of $f^*$ is not worse than that of $f$. For these proofs, we use $p^*$ to denote the probabilities referred to the case where the LPPM used is $f^*$. Also, we use $\mathbf{z}^{-s}\doteq[z^1,z^2,\cdots,z^{s-1},z^{s+1},\cdots, z^r]$.

Our goal is to prove that $\min_h\PAE(f,h,r,s)\leq \min_h\PAE(f^*,h,r,s)$, i.e., that $f^*$ does not achieve less privacy than $f$ against an optimal adversary that minimizes $\PAE$:
\begin{align*}
 &\min_h\PAE(f,h,r,s) \\
 \quad&=\sum_{\mathbf{z}^{r}\in\mathcal{Z}^{r}} \min_{\hat{x}^s} \left[ \sum_{x^s\in\mathcal{X}} \pi(x^s) p(\mathbf{z}^r|x^s) d_P(x^s,\hat{x}^s)\right]\\
 \quad  &\stackrel{(a)}{\leq}\sum_{z^s\in\mathcal{Z}} \min_{\hat{x}^s} \left[\sum_{\substack{\mathbf{z}^{-s}\\ \in\mathcal{Z}^{-s}}} \sum_{x^s\in\mathcal{X}} \pi(x^s) p(\mathbf{z}^r|x^s) d_P(x^s,\hat{x}^s)\right]\\
 \quad  &=\sum_{z^s\in\mathcal{Z}} \min_{\hat{x}^s} \left[ \sum_{x^s\in\mathcal{X}} \pi(x^s) p(z^s|x^s) d_P(x^s,\hat{x}^s) \right]\\
 \quad  &=\sum_{z^s\in\mathcal{Z}} \min_{\hat{x}^s} \left[ \sum_{x^s\in\mathcal{X}} \pi(x^s) f^*(z^s|x^s) d_P(x^s,\hat{x}^s) \right]\\
 \quad  &\stackrel{(b)}{=}\sum_{\mathbf{z}^r\in\mathcal{Z}^r} \min_{\hat{x}^s} \left[ \sum_{x^s\in\mathcal{X}} \pi(x^s) p(\mathbf{z}^{-s}) f^*(z^s|x^s) d_P(x^s,\hat{x}^s) \right]\\
 \quad  &=\min_h\PAE(f^*,r,s)\,.
\end{align*}
Step $(a)$ comes from splitting the summation over $\mathbf{z}^r$ into two summations: one over $z^s$ and the other over the complement. Then, computing the summation (over $\mathbf{z}^{-s}$) of the minima over $\hat{x}^s$ is smaller or equal than computing the minimum of the summation. Step $(b)$ follows from the fact that $\mathbf{z}^{-s}$ is independent of $z^s$ and $x^s$ in the hardwired model and with a memoryless LPPM $f^*$.

\subsection{Convergence of the EM sequence to the MLE of the mobility profile.}
\label{sec:proofEM}

We prove the convergence of the EM iteration in \eqref{eq:updatepi} to the maximum likelihood estimator of the mobility profile, \emph{for memoryless and output-based LPPMs} only. Let $\mathcal{P}$ be the probability simplex, i.e., the set of valid mobility profiles $\mathcal{P}\doteq\{\pi|\sum_{i=1}^{|\mathcal{X}|} \pi_i=0,\quad \pi_i\geq 0\}$. Then, the MLE is
\begin{equation}
 \hat{\pi}^r_{ML}= \underset{\pi\in\mathcal{P}}{\text{ argmax}} \log p(\mathbf{z}^r|\pi)\,.
\end{equation}

In~\cite{wu1983convergence,agrawal2001design}, authors show that if the likelihood function (i.e., $\log p(\mathbf{z}^r|\pi)$) has a unique global maximum over $\mathcal{P}$ and the derivatives $\partial Q(\pi,\pi^t)/\partial \pi$ are continuous over $\pi$ and $\pi^t$, then any EM sequence $\{\pi^0,\pi^1,\pi^2,\cdots\}$ computed as in \eqref{eq:updatepi} converges to the unique global maximum $\hat{\pi}_{ML}^r$. We now prove that our problem meets these requirements, and refer to \cite{wu1983convergence,agrawal2001design} for the complete details of the proof.

First, we prove that $\log p(\mathbf{z}^r|\pi)$ is strictly concave and has a unique global maximum over $\mathcal{P}$. By definition, it is easy to see that $\mathcal{P}$ is convex, i.e., given two profiles $\pi,\pi'\in\mathcal{P}$, we can check that $\pi''\doteq \lambda \pi + (1-\lambda) \pi' \in\mathcal{P}$ for $\lambda\in[0, 1]$. On the other hand, we can write $\log p(\mathbf{z}^r|\pi)=\sum_{s=1}^r \log p(z^s|\mathbf{z}^{s-1},\pi)$ and show that
\begin{equation}
  p(z^s|\mathbf{z}^{s-1},\pi)=\sum_{i=1}^{|\mathcal{X}|} f(z^s|\mathbf{z}^{s-1},x^s=x_i)\cdot \pi_i\,,
\end{equation}
where $f(z^s|\mathbf{z}^{s-1},x^s=x_i)$ is given by the LPPM (it does not require $\pi$ for its computation, since it is an output-based LPPM). This means that $p(z^s|\mathbf{z}^{s-1},\pi)$ is linear with $\pi$, and therefore $\log p(z^s|\mathbf{z}^{s-1},\pi)$ is strictly concave. This implies that $\log p(\mathbf{z}^r|\pi)$ is also strictly concave, since it is the sum of strictly concave functions. Since $\mathcal{P}$ is a convex set, then $\log p(\mathbf{z}^r|\pi)$ has a unique global maximum over $\mathcal{P}$.

On the other hand, it is easy to see that the derivatives $\partial Q(\pi,\pi^t)/\partial \pi$ are continuous in $\pi$ and $\pi^t$ (note that $\pi_i\in[0,1]$), which concludes the proof. 

The proof for memoryless LPPMs is the same, since they are a sub-type of output-based LPPMs. This result is not true, however, for full-type LPPMs, since $p(z^s|\mathbf{z}^{s-1},x^s=x_i)$ depends on the mobility profile (through all the other inputs).

\subsection{Experiment MK: training with data from other users.}
\label{sec:appMK}

\begin{figure} 
  \centering
  \subfloat[TaxiCab, $\Expo$.]{\includegraphics[width=0.49\linewidth]{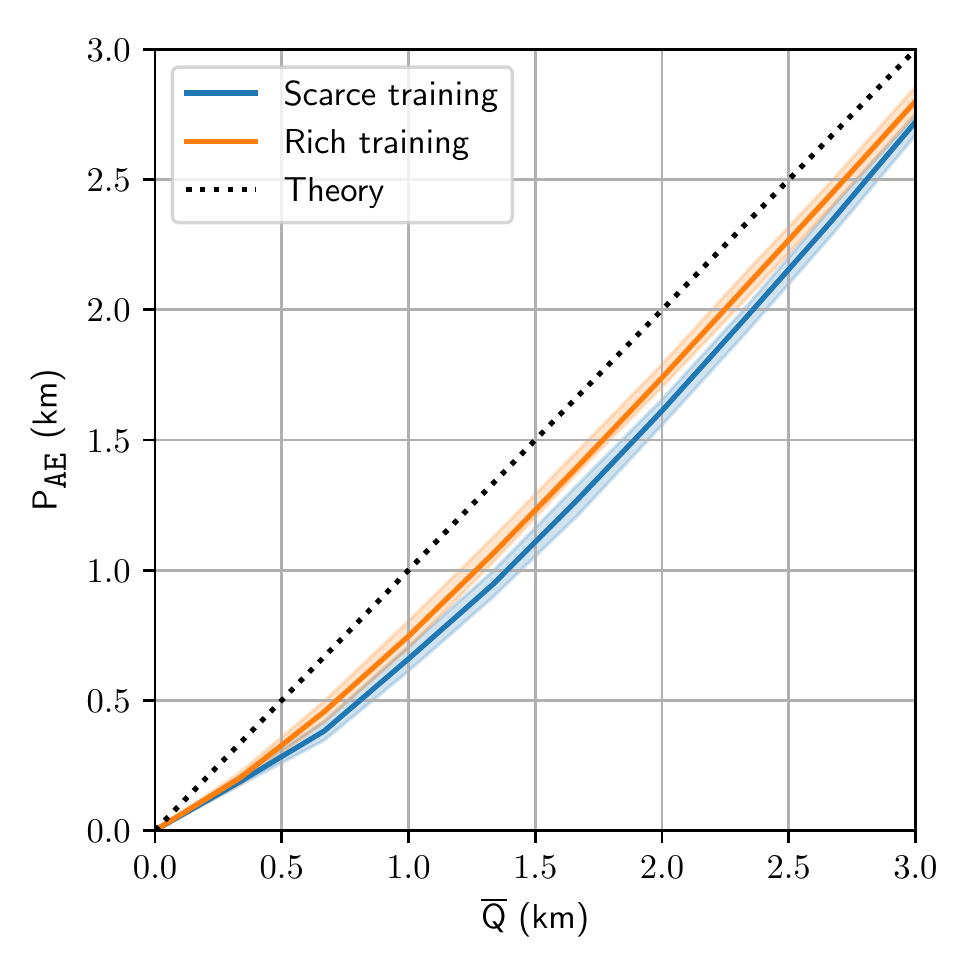}} 
  \hfil
  \subfloat[TaxiCab, $\LH$.]{\includegraphics[width=0.49\linewidth]{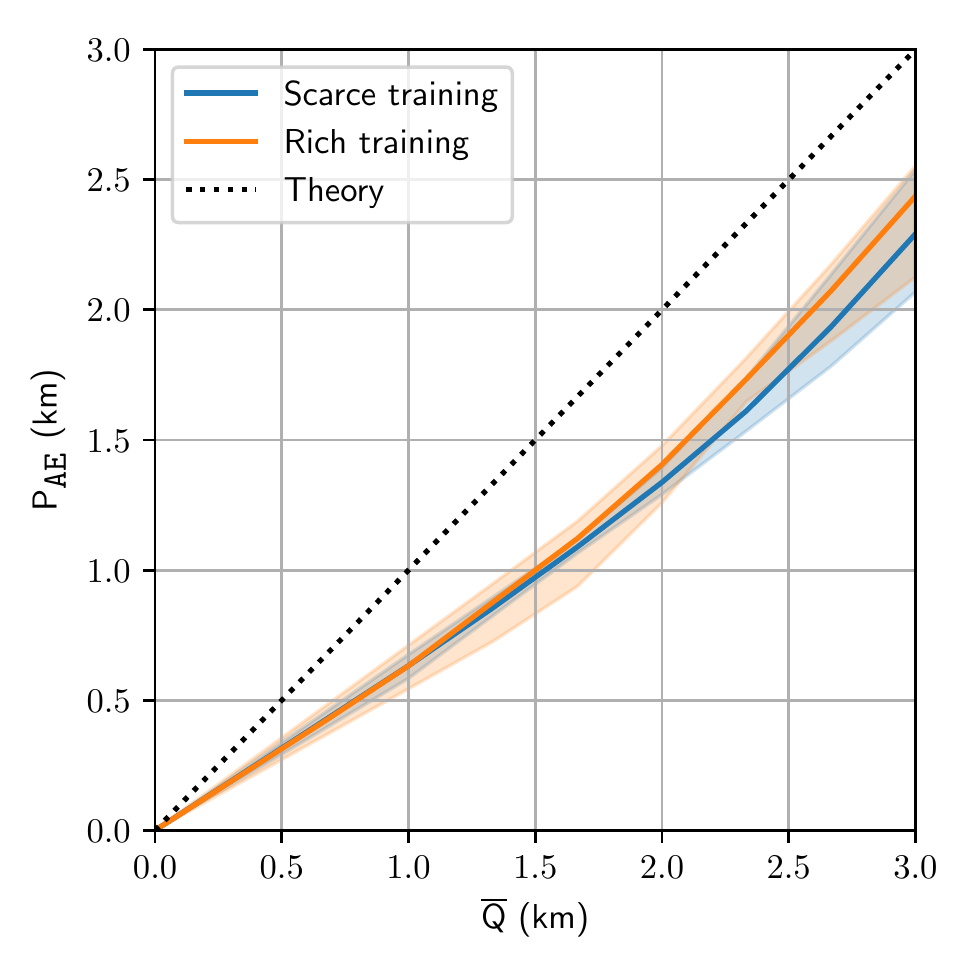}} 
  \caption{Experiment MK: Performance of $\Expomk$ and $\LHmk$ against the optimal Markov attack in Taxicab dataset, when the LPPMs are trained with data \emph{from other users.}}
  \label{fig:Eval1-Exp22_2}
\end{figure}

We repeat the Experiment MK described in Sect.~\ref{sec:experimentMK}, but we train each user's LPPM using past days \emph{of a different user}. Fig.~\ref{fig:Eval1-Exp22_2} shows the results of these experiments. Here, \emph{scarce training} and \emph{rich training} refer to the case where we train the LPPM of each user with the past day or the past seven days of a different user. The performance of $\Expomk$ and $\LHmk$ is slightly worse than in Fig.~\ref{fig:Eval1-Exp22}, i.e., the LPPMs achieve less privacy for the same quality loss, but it is nonetheless very similar. This supports our hypothesis that road restrictions (and not personal user preferences) are the main source of correlation between locations in this dataset.

\section*{Acknowledgments}

This work is partially supported by the Agencia Estatal de Investigaci{\'o}n (Spain) and the European Regional Development Fund (ERDF) under project WINTER (TEC2016-76409-C2-2-R), and by the Xunta de Galicia and the ERDF under projects Agrupacion Estratexica Consolidada de Galicia accreditation 2016-2019, Grupo de Referencia ED431C2017/53, and Red Tematica RedTEIC 2017-2018. Simon Oya is funded by the Spanish Ministry of Education, Culture and Sport under the FPU grant.

\bibliographystyle{IEEEtran}
\bibliography{bibliolocation}

\end{document}